\pdfoutput=1

\documentclass[aps,prd,onecolumn,groupedaddress,preprintnumbers,superscriptaddress,nofootinbib]{revtex4}
\usepackage{amsmath,amssymb,latexsym}
\usepackage{graphicx}
\def\lsim{\mathrel{\rlap{\lower4pt\hbox{\hskip1pt$\sim$}}
    \raise1pt\hbox{$<$}}}                

\begin{document}

\title{
Non-Gaussianity from Lifshitz Scalar
}


\author{
Keisuke Izumi
}\email[]{keisuke.izumi@ipmu.jp}
\affiliation{
Institute for the Physics and Mathematics of the Universe (IPMU), The
University of Tokyo, 5-1-5 Kashiwanoha, Kashiwa, Chiba 277-8583, Japan
}
\author{
Takeshi Kobayashi
}\email[]{takeshi.kobayashi@ipmu.jp}
\affiliation{
Institute for Cosmic Ray Research, The University of Tokyo, 5-1-5
Kashiwanoha, Kashiwa, Chiba 277-8582, Japan 
}
\author{
Shinji Mukohyama
}\email[]{shinji.mukohyama@ipmu.jp}
\affiliation{
Institute for the Physics and Mathematics of the Universe (IPMU), The
University of Tokyo, 5-1-5 Kashiwanoha, Kashiwa, Chiba 277-8583, Japan
}

\preprint{IPMU10-0133}
\preprint{ICRR-Report-571} 


\begin{abstract}
 A Lifshitz scalar with the dynamical critical exponent $z\! =\! 3$
 obtains scale-invariant, super-horizon field fluctuations without the
 need of an inflationary era. Since this mechanism is due to the special
 scaling of the Lifshitz scalar and persists in the presence of
 unsuppressed self-couplings, the resulting fluctuation spectrum can
 deviate from a Gaussian distribution. We study the non-Gaussian nature
 of the Lifshitz scalar's intrinsic field fluctuations, and show that
 primordial curvature perturbations sourced from such field fluctuations
 can have large non-Gaussianity of order $f_{\mathrm{NL}} = O(100)$,
 which will be detected by upcoming CMB observations. We compute the
 bispectrum and trispectrum of the fluctuations, and discuss their
 configurations in momentum space. In particular, the bispectrum is
 found to take various shapes, including the local, equilateral, and
 orthogonal shapes. Intriguingly, all integrals in the in-in formalism
 can be performed analytically. 
\end{abstract}

\maketitle

\section{introduction}

Ho\v{r}ava-Lifshitz gravity~\cite{Horava:2009uw} is attracting much
attention as one of candidates for the theory of quantum gravity because
of its power-counting renormalizability, which is realized by the
Lifshitz scaling 
\begin{eqnarray}
\vec{x} \to b \vec{x}, \qquad t \to b^z t, 
\label{scaling}
\end{eqnarray}
with the dynamical critical exponent $z\geq 3$ in the ultraviolet (UV). 
There are many attempts to investigate properties and implications of
this theory~\cite{Mukohyama:2009gg,others}.

It is natural to suppose that not only gravitational fields but also 
other fields exhibit the same Lifshitz scaling in the UV. Even if they
classically have different scalings, quantum corrections should render
them to have the same scaling. A Lifshitz scalar with $z=3$ can obtain
scale-invariant, super-horizon field fluctuations even without
inflation~\cite{Mukohyama:2009gg}, thus can source the primordial
curvature perturbations through mechanisms such as the curvaton
scenario~\cite{Lyth:2001nq} or the modulated
decay~\cite{Dvali:2003em}. It is noteworthy that this value of $z$ is
the minimal value for which gravity is power-counting renormalizable.

In order to discern this production mechanism of the primordial
perturbation from others, we need to investigate distinct features in
observables such as the cosmic microwave background. In this respect,
non-Gaussianity has been considered as one of the promising approaches
to distinguish production mechanisms. For this reason, there are
on-going efforts to detect or constrain non-Gaussian nature of the
primordial perturbation~\cite{Komatsu:2010fb}. Towards identification of the
production mechanism by future observations, theoretical analyses of
non-Gaussianity in various cosmological scenarios have been
performed~\cite{Maldacena:2002vr,nongaussianity-1,nongaussianity-2,nongaussianity-3}.

In this paper, we focus on primordial non-Gaussianity from a Lifshitz
scalar and calculate its bispectrum and trispectrum. With the dynamical 
critical exponent $z=3$, the scaling dimension of the Lifshitz scalar is
zero and, thus, nonlinear terms in the action are unsuppressed unless
forbidden by symmetry or driven to small values by renormalization. It
is those nonlinear terms that we expect to produce non-Gaussianity. 
Even when the Lifshitz scalar's field fluctuations are linearly
transformed to the curvature perturbations (which can be realized by the
curvaton mechanism or/and modulated decay), it turns out that the
produced bispectrum can be large enough to be observed in future
observations. We find three independent cubic terms dominant in the UV,
each of which gives different shape dependence of the
bispectrum. Roughly speaking, they correspond to local, equilateral and
orthogonal shapes, respectively.

The rest of this paper is organized as follows. In section
\ref{sec:review} we review generation of scale-invariant cosmological
perturbations from a Lifshitz scalar. In section \ref{sec:estimation} we
estimate the size of non-Gaussianity and see that the nonlinear
parameter $f_{\mathrm{NL}}$ can be as large as $O(100)$. In section
\ref{sec:shape} we concretely show the momentum dependence of the bispectrum
and trispectrum. Section \ref{sec:summary} is devoted to a summary of
this paper and discussions. In appendix \ref{appendix} we derive 
the set of independent cubic and quartic terms dominant in the UV.

\section{Scale-invariant power spectrum from Lifshitz scalar}
\label{sec:review}

In this section, we review the mechanism for generation of 
scale-invariant cosmological perturbations from a Lifshitz
scalar~\cite{Mukohyama:2009gg}. 
The action for a Lifshitz scalar $\Phi$ in Minkowski background is 
\begin{eqnarray}
S_\Phi = \frac{1}{2} \int dt d^3x\ 
\left[ (\partial _t \Phi)^2 + \Phi {\tilde{\cal O}}\Phi
+O(\Phi^3)
\right],
\label{LifshitzScalarMinkowski}
\end{eqnarray}
where
\begin{eqnarray}
{\tilde{\cal O}} = (-1)^{z+1} \frac{1}{\tilde M^{2(z-1)} } \Delta^z 
 +(-1)^z \frac{s_{z-1}}{\tilde M^{2(z-2)} } \Delta^{z-1}
+ \cdots +s_{1}\Delta - {\tilde m}^2.
\label{actionPhi}
\end{eqnarray}
$\Delta = \delta^{ij}\partial_i \partial_j$, 
${\tilde M}$ and ${\tilde m}$ are mass scales and $s_n$
are dimensionless constants. Here, it is supposed that the time kinetic
term is already canonically normalized, and thus nonlinear terms in the
action indicated by $ O(\Phi^3)$ do not include time derivatives. On 
the other hand, those nonlinear terms can include spatial
derivatives. 
Also, the sign of the first term in the right hand side of
(\ref{actionPhi}) is set by requiring stability in the UV.

In the UV, the first term in ${\tilde{\cal O}}$ is dominant and the
field $\Phi$ described by the action (\ref{actionPhi}) exhibits the
Lifshitz scaling (\ref{scaling}) with 
\begin{eqnarray}
\Phi \to b^{(z-3)/2} \Phi.
 \label{eqn:Lifshitz-scaling}
\end{eqnarray}
We find that for $z=3$, the scaling dimension of $\Phi$ is zero and 
thus the amplitude of quantum fluctuations of $\Phi$ is expected to be
independent of the energy scale of the system of interest. This
indicates that the power spectrum of quantum fluctuations of $\Phi$
in an expanding universe should be scale-invariant. Intriguingly, the
minimal value of $z$ for which Ho\v{r}ava-Lifshitz gravity is
power-counting renormalizable is also $3$. Hereafter, we consider the
$z=3$ case.

Now let us consider the Lifshitz scalar $\Phi$, specialized to the case
with $z=3$, in a flat FRW background
\begin{eqnarray}
ds^2= -dt+a(t)^2\delta_{ij}dx^i dx^j,
\end{eqnarray}
to investigate generation of cosmological perturbations. We just need to 
replace the volume element $d^3x$ by $d^3xa(t)^3$ and the spatial
Laplacian $\Delta$ by $\Delta/a(t)^2$ in the action
(\ref{LifshitzScalarMinkowski}) with $z=3$. We expand the scalar field
$\Phi$ around a homogeneous v.e.v. $\Phi_0$ as $\Phi=\Phi_0+\phi$. 
Throughout this paper we consider the UV regime in which the Hubble
expansion rate $H$ is much higher than mass scales in the scalar field
action. In this regime, the Hubble friction is so strong that the time
dependence of the background $\Phi_0$ is unimportant. For this reason,
hereafter, we treat $\Phi_0$ as a constant. The action for the
perturbation $\phi$ is then written as 
\begin{eqnarray}
S_\phi = \frac{1}{2} \int dt d^3x \ a(t)^3 
\left[ (\partial _t \phi)^2 + \phi {\cal O}\phi
+O(\phi^3)
\right],
\end{eqnarray}
where
\begin{eqnarray}
{\cal O} = \frac{1}{ M^4a(t)^6}\Delta^3 
-\frac{ s}{ M^2a(t)^4}\Delta^2
+\frac{c_s^2}{a(t)^2}\Delta- m^2.
\end{eqnarray}
$M$ and $m$ are mass scales and $s$ and $c_s^2$ are dimensionless
constants. In the UV, the quadratic action for $\phi$ is simply
\begin{equation}
 S_2 = \frac{1}{2} \int dt d^3x\,  a(t)^3 \left\{ (\partial_t \phi)^2 +
				     \frac{1}{M^4 a(t)^6} \phi \Delta ^3
				     \phi \right\}.
 \label{S2}
\end{equation}

As discussed after (\ref{eqn:Lifshitz-scaling}), the scaling dimension
of $\Phi$ and thus $\phi$ is zero, 
\begin{eqnarray}
\phi\to b^0\phi, \label{scalingphi}
\end{eqnarray}
and its power-spectrum should be scale-invariant. Since $\phi$ is
scale-invariant and there is only one scale $M$ in the UV quadratic
action (\ref{S2}), we expect that the power-spectrum should be roughly 
\begin{eqnarray}
  \left\langle \phi\phi \right\rangle \sim M^2. 
   \label{phiphi-roughly}
\end{eqnarray}

\begin{figure}[t]
  \begin{center}
   \includegraphics[keepaspectratio=true,height=100mm]{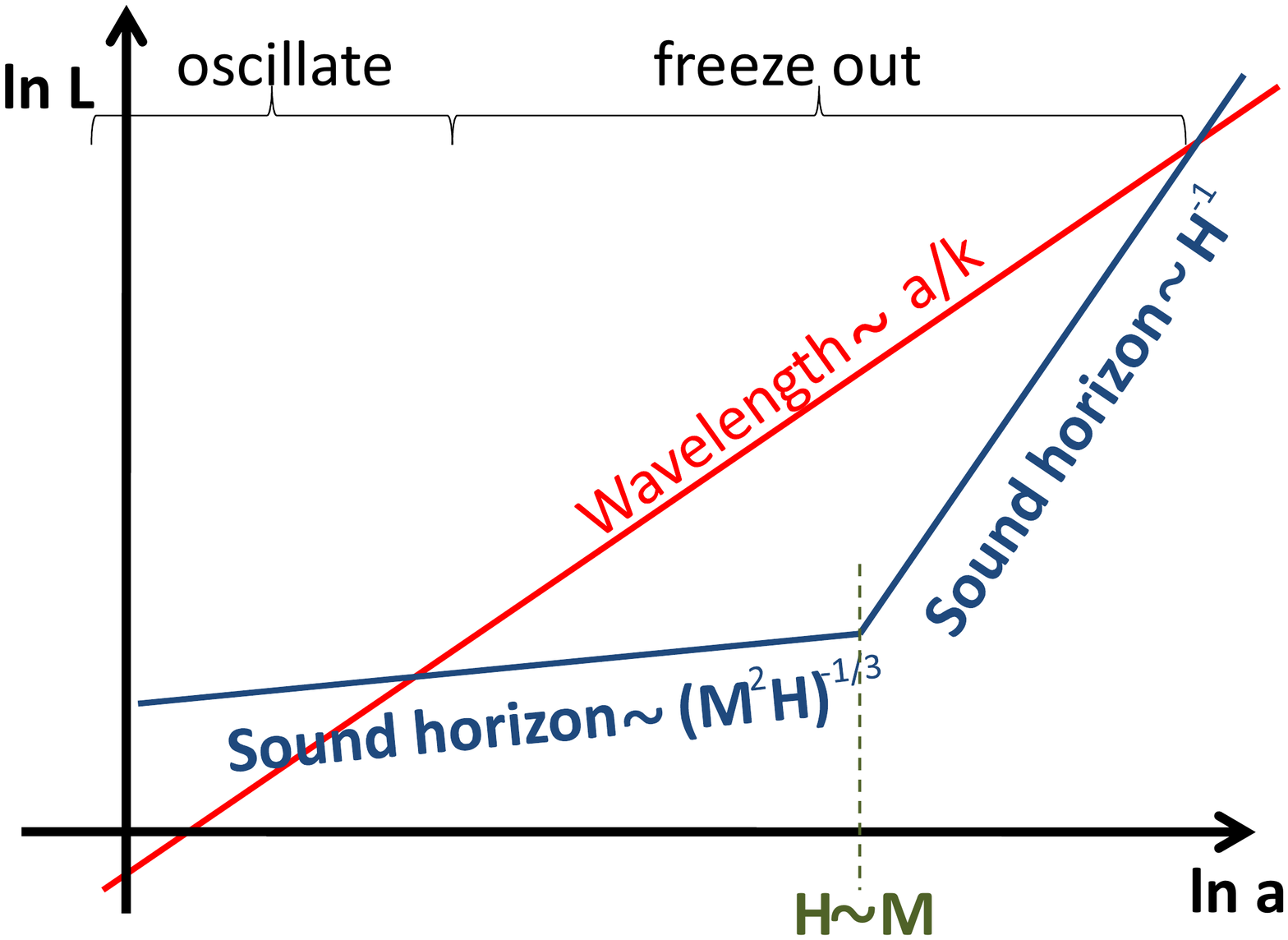}
  \end{center}
  \caption{A schematic picture of the evolution of cosmological
 perturbations for a power-law expansion $a\propto t^p$ with $1/3<p<1$. 
 The physical wavelength exits the sound horizon $\sim (M^2H)^{-1/3}$ in
 the UV and re-enters the sound horizon $\sim H^{-1}$ in the IR. We
 suppose that the scale-invariant perturbations of the Lifshitz scalar
 are converted into those of radiation by curvaton mechanism or/and
 modulated decay when physical wavelengths of interest are outside the
 sound horizon. Thus, strictly speaking, the sound horizon in the UV is
 for the Lifshitz scalar and that in the IR is for radiation.} 
    \label{fig:wave.pdf}
\end{figure}

Now let us calculate the power spectrum concretely. By solving the 
Heisenberg equation obtained from the quadratic action (\ref{S2}),
operator $\phi$ can be expanded as
\begin{equation}
 \phi (t, \boldsymbol{x}) =\frac{1}{(2 \pi)^3 } \int d^3 k \, 
 e^{i \boldsymbol{k} \cdot \boldsymbol{x}} \phi_{\boldsymbol{k}}(t),
\end{equation}
and
\begin{equation}
 \phi_{\boldsymbol{k}}(t) = u_{\boldsymbol{k}}(t) a_{\boldsymbol{k}} +
  u_{-\boldsymbol{k}}^* (t) a_{-\boldsymbol{k}}^{\dagger},
  \label{phi}
\end{equation}
where
\begin{equation}
 u_{\boldsymbol{k}}(t) = \frac{M}{2^{1/2} k^{3/2} }
 \exp\left(-i \frac{k^3 }{M^2} \int^t \frac{dt'}{a(t')^3} \right).
 \label{modefunction}
\end{equation}
The mode function $u_{\boldsymbol{k}}(t)$ is chosen so that 
its asymptotic behavior in the Minkowski limit ($a(t)\to\ $const.) is the same as 
the positive-frequency mode function in Minkowski background. The vacuum
state $\left|0\right\rangle$ is defined as usual by 
\begin{eqnarray}
a_{\boldsymbol{k}}\left|0\right\rangle=0.
\end{eqnarray} 
The mode function $u_{\boldsymbol{k}}(t)$ approaches a constant value in
the $a(t)\to \infty$ limit if the integral $\int^{t_\infty} dt / a(t)^3$
converges, where $t_\infty$ is the time corresponding to 
$a(t)\to \infty$. The power-law expansion $a(t) \propto t^p$ with
$p>1/3$ satisfies this condition. Under this condition, when the
physical wavelength $a(t)/k$ becomes as long as the size of the sound
horizon $\sim (M^2H)^{-1/3}$, the mode function $u_{\boldsymbol{k}}(t)$
stops oscillating and freezes out. (See in Fig.(\ref{fig:wave.pdf}).)
Note that the physical wavelength at sound horizon exit is super-horizon
size in the UV, i.e. when $H \gg M$. 

The commutation relations of operators are defined in the usual manner
as 
\begin{eqnarray}
[\phi(\boldsymbol{x}), \pi(\boldsymbol{y})]=i\delta^3(\boldsymbol{x}-\boldsymbol{y}),
\qquad
[\phi(\boldsymbol{x}), \phi(\boldsymbol{y})]=
[\pi(\boldsymbol{x}), \pi(\boldsymbol{y})]
=0,
\end{eqnarray}
where
\begin{eqnarray}
\pi(\boldsymbol{x}) = \frac{\delta S_2}{\delta \partial_t \phi(\boldsymbol{x})}
=a(t)^3 \partial_t \phi(\boldsymbol{x}),
\end{eqnarray}
and hence the operators $a_{\boldsymbol{k}}$ and $a_{\boldsymbol{k}}^\dagger$ 
satisfy 
\begin{eqnarray}
[a_{\boldsymbol{k}},a_{\boldsymbol{k}'}^\dagger]=
(2\pi)^3\delta^3(\boldsymbol{k}-\boldsymbol{k}') ,
\qquad
[a_{\boldsymbol{k}},a_{\boldsymbol{k}'}]=
[a_{\boldsymbol{k}}^\dagger,a_{\boldsymbol{k}'}^\dagger]=0.
\end{eqnarray}
Power spectrum ${\cal P}_\phi$ is defined as 
\begin{eqnarray}
\left\langle 0 | \phi_{\boldsymbol{k}}\phi_{\boldsymbol{k} '} | 0 \right\rangle
=(2\pi)^3 \delta^3(\boldsymbol{k}+\boldsymbol{k}') P_\phi, 
\qquad
P_\phi = \frac{2\pi^2}{k^3} {\cal P}_\phi,
\label{power}
\end{eqnarray}
so that
\begin{eqnarray}
\left\langle 0 | \phi^2 | 0 \right\rangle = \int \frac{dk}{k} {\cal P}_\phi.
\end{eqnarray}
Substituting eqs.(\ref{phi}) and (\ref{modefunction}) into
eq.(\ref{power}),  we obtain 
\begin{equation}
 {\cal{P}}_{\phi} = \frac{M^2}{(2 \pi)^2}. 
\end{equation}

In this paper, we suppose that the primordial fluctuations of the
Lifshitz scalar field $\phi$ are almost linearly transformed to the
curvature perturbations and that large non-Gaussianity is not
created in the transformation itself. In scenarios like the curvaton
mechanism~\cite{Lyth:2001nq} and the modulated
decay~\cite{Dvali:2003em}, this assumption can be realized naturally,
while violation of the assumption is also possible in these scenarios. 
With the assumption, the curvature perturbation $\zeta$ is related to
the field fluctuation $\phi$ as 
\begin{eqnarray}
\zeta =\frac{\phi}{\mu} + O(1)\times \left(\frac{\phi}{\mu}\right)^2
 + O(1)\times \left(\frac{\phi}{\mu}\right)^3 + \cdots,
 \label{phi-to-zeta}
\end{eqnarray}
where $\mu$ is an energy scale. Therefore, the power spectrum of the 
curvature perturbation takes over the scale-invariance as
\begin{eqnarray}
{\cal{P}}_{\zeta}
 =\mu^{-2}{\cal{P}}_{\phi} = \frac{M^2}{(2 \pi)^2 \mu^2}. 
\end{eqnarray}
The COBE normalization~\cite{Komatsu:2010fb} sets 
${\cal{P}}_{\zeta}^{1/2} \simeq 4.9 \times 10^{-5}$,  
and thus 
\begin{eqnarray}
\frac{M}{\mu} \simeq 3.0 \times 10^{-4}. 
 \label{COBEnormalization}
\end{eqnarray}

\section{Order estimate for $f_{\mathrm{NL}}$}
\label{sec:estimation}

Let us move on to discussions about non-Gaussianity. In the previous 
section we have seen that a free Lifshitz scalar with the dynamical
critical exponent $z=3$ can produce scale-invariant perturbations. The
essential reason for this is that the scaling dimension of the scalar is
zero at high energy. This implies that at high energy, some nonlinear
operators could be as important as leading operators in the quadratic
action and that large non-Gaussianity could be generated by those
nonlinear operators. On the other hand, because of power-counting
renormalizability, those nonlinear operators do not completely dominate
over the quadratic terms and thus do not spoil the analysis of scaling
dimensions and the scale-invariance of the power-spectrum unless the
theory gets really strongly coupled. In this section, we shall present
order estimates for the bispectrum of curvature perturbations and the
corresponding nonlinear parameter $f_{\mathrm{NL}}$, deferring detailed
calculations until the next section.

As presented in (\ref{phi-to-zeta}), throughout this paper we assume
that perturbations of the Lifshitz scalar are almost linearly
transformed to curvature perturbations. In our calculation of the bispectrum 
and trispectrum of curvature perturbations, we thus ignore nonlinear
terms in the right hand side of (\ref{phi-to-zeta}) and take into
account the linear term only. In particular,
\begin{equation}
\left\langle \zeta\zeta \right\rangle \simeq  
 \mu^{-2}\left\langle \phi\phi \right\rangle, \quad
\left\langle \zeta\zeta\zeta \right\rangle \simeq
 \mu^{-3}\left\langle \phi\phi\phi \right\rangle. 
 \label{phiphi-to-zetazeta}
\end{equation} 
This treatment is justified by the fact that perturbations of the
Lifshitz scalar have large non-Gaussianity. Actually, the bispectrum and
trispectrum would obtain corrections from nonlinear terms in
(\ref{phi-to-zeta}), but those corrections would be smaller than large
contributions from the Lifshitz scalar's intrinsic non-Gaussianity. In
terms of the nonlinear parameter $f_{\mathrm{NL}}$, corrections due to nonlinear
terms in (\ref{phi-to-zeta}) are typically $O(1)$ but we shall see below
that the scalar field's intrinsic non-Gaussianity can lead to
$f_{\mathrm{NL}}=O(100)$.

We shall adopt the so called in-in
formalism~\cite{Maldacena:2002vr,Weinberg:2005vy} to calculate the
bispectrum and trispectrum of the Lifshitz scalar. The leading
contribution to the bispectrum is given by the following formula (see the
next section for details)
\begin{equation}
 \left\langle \phi\phi\phi \right\rangle = 
 i   \left\langle \left[\int dt H_{3} ,
 \phi\phi\phi  \right] \right\rangle ,
 \label{in-in-bispectrum}
\end{equation} 
where $H_3$ represents cubic terms in the interaction
Hamiltonian. Dominant terms in $\int dt H_{3}$ are marginal ones,
i.e. those terms whose scaling dimensions are zero. Actually, as shown
in Appendix~\ref{appendix}, there are three (and only three) independent
marginal cubic operators in the action in the UV: 
\begin{eqnarray}
 S_3 = \int dt d^3x \, \frac{1}{M^5 a(t)^3} \left\{ \alpha_1 \phi^2 \Delta^3
  \phi + \alpha_2  (\Delta^2 \phi) (\partial_i \phi)^2
 + \alpha_3     (\Delta \phi)^3\right\}, 
 \label{S3}
\end{eqnarray}
where $\alpha_i$ are dimensionless parameters. (The fist term can be
forbidden by the shift symmetry if one likes.) Evidently, validity of
perturbative expansion (in the in-in formalism) requires $\alpha_i$ be 
smaller than unity.  The corresponding cubic operators in the
interaction Hamiltonian are 
\begin{eqnarray}
 H_3(t) = -\int d^3x \, \frac{1}{M^5 a^3} \left\{ \alpha_1 \phi^2
    \Delta^3  \phi + \alpha_2  (\Delta^2 \phi) (\partial_i \phi)^2
 + \alpha_3     (\Delta \phi)^3\right\}. 
 \label{H3}
\end{eqnarray}
Each of these dominant cubic terms includes six spatial
derivatives and gives zero scaling dimension to $\int dt H_{3}$. 
Combining this with the fact that the scaling dimension of $\phi$ is
zero, we conclude that the bispectrum of $\phi$ given by
(\ref{in-in-bispectrum}) should be scale-independent, and thus
\begin{equation}
  \left\langle \phi\phi\phi \right\rangle \sim \alpha M^3,
   \label{phiphiphi-roughly}
\end{equation}
where $\alpha$ stands for the most dominant one among $\alpha_i$
($i=1,2,3$).

Roughly speaking, the non-linear parameter $f_{\mathrm{NL}}$ is defined so that 
\begin{equation}
f_{\mathrm{NL}} \sim \frac{\left\langle \zeta\zeta\zeta \right\rangle}
{\left\langle\zeta\zeta\right\rangle^2}.
\end{equation}
Thus, combining this with (\ref{phiphi-to-zetazeta}),
(\ref{phiphi-roughly}) and (\ref{phiphiphi-roughly}), we obtain
\begin{equation}
f_{\mathrm{NL}} \sim \alpha \left(\frac{M}{\mu}\right)^{-1} \sim 3
 \times 10^3 \alpha .  
\label{eq:estimation}
\end{equation} 
Here, we have used the COBE normalization (\ref{COBEnormalization}). 
In the next section, we obtain a more precise expression for
$f_{\mathrm{NL}}$ in terms of $\alpha_i$.

As already stated, validity of perturbative expansion requires that
the dimensionless parameters $\alpha_i$ be smaller than unity. We find
from the order estimate (\ref{eq:estimation}) that $f_{\mathrm{NL}}$ can be
large, e.g. as large as $O(100)$, even if $\alpha_i$ are reasonably
small.

\section{Shapes of Non-Gaussianities}
\label{sec:shape}

We have seen that the Lifshitz scalar's non-Gaussian intrinsic
fluctuations can leave observable non-Gaussianities in the sky.
In this section, we compute the bispectrum and trispectrum of Lifshitz
scalar fluctuations and discuss their shapes. Since a Lifshitz scalar
with $z=3$ can have all possible self-coupling terms containing six
spatial derivatives (cf. (\ref{S3}), (\ref{S4})), the resulting
correlation functions can take various configurations in momentum
space. We will see especially that the generated bispectrum includes
local, equilateral, and orthogonal shapes. 
We also discuss non-Gaussianity of the primordial curvature
perturbations sourced by the Lifshitz scalar's field fluctuations. By
comparing with the latest observational constraints, we obtain bounds on
the Lifshitz scalar's self-coupling strengths.

\subsection{Bispectrum}
\label{subsec:bispectrum}

We make use of the prescription of the in-in formalism (see
e.g. \cite{Maldacena:2002vr,Weinberg:2005vy}) for calculating
expectation values of a product~$Q(t)$ of field operators at 
time~$t$, 
\begin{equation}
 \left\langle  Q(t)  \right\rangle
 = \left\langle  \left[ \overline{T} \exp \left( i
   \int^t_{t_0} H_I (t')dt'
\right)  \right] Q^I(t)
 \left[ T \exp \left(- i
   \int^t_{t_0} H_I (t')dt'
\right)  \right]
\right\rangle, \label{inin}
\end{equation} 
where $T$ denotes time-ordering, $\overline{T}$ is anti-time-ordering,
$Q^I(t)$ is the product~$Q(t)$ in the interaction picture, and $H_I(t)$
is the interaction part of the Hamiltonian in the interaction picture.  
We take the time~$t_0$ to be at very early times when the fluctuation
wavelengths are well inside the sound horizon. 
In the present case, introducing the time~$d \tau = dt/a^3$, then the integral
can be taken in terms of $\tau$ from $- \infty$ to some time after the
sound horizon exit.\footnote{Note that the lower limit of integration in
(\ref{inin}) is shifted to $-\infty (1 - i \epsilon)$ on the right side
of $Q^I$, and to $-\infty (1 + i \epsilon)$ on the left side, so that
the oscillating exponents become exponentially decreasing. This
prescription corresponds to picking up the vacuum state.}

In order to obtain the bispectrum of the Lifshitz scalar fluctuations,
we expand (\ref{inin}) in terms of $H_I$ and compute terms which are
first order in the 3-point interaction Hamiltonian~(\ref{H3}). We then find
\begin{align}
 \left\langle \phi_{\boldsymbol{k_1}} (t) \phi_{\boldsymbol{k_2}} (t)
   \phi_{\boldsymbol{k_3}} (t) \right\rangle & = 
 i \int^{t}_{t_0} dt' \langle \left[ H_{3} (t'),
 \phi_{\boldsymbol{k_1}} (t) \phi_{\boldsymbol{k_2}} (t) 
   \phi_{\boldsymbol{k_3}} (t)  \right] \rangle \\
 & = -i (2 \pi)^3 \delta^{(3)}
 (\boldsymbol{k_1} + \boldsymbol{k_2} + \boldsymbol{k_3} )
 f(\boldsymbol{k_1}, \boldsymbol{k_2}) \\
 & \qquad \qquad
 \times
 u_{\boldsymbol{k_1}}(t) u_{\boldsymbol{k_2}}(t)
 u_{\boldsymbol{k_3}}(t) 
\int^t_{t_0}  \frac{dt'}{M^5 a(t')^3} 
 u_{-\boldsymbol{k_1}}^*(t') u_{-\boldsymbol{k_2}}^*(t')
 u_{-\boldsymbol{k_3}}^*(t')  + \mathrm{c.c.} \\
 & = - \frac{(2\pi)^3 M^3}{2^2} \frac{ \delta^{(3)} (\boldsymbol{k_1} +
 \boldsymbol{k_2} + \boldsymbol{k_3} )  f(\boldsymbol{k_1},
 \boldsymbol{k_2})  }{(k_1 k_2 k_3)^3 (k_1^3 +
 k_2^3 +  k_3^3)}, \label{bispectrum}
\end{align}
where we have defined
\begin{equation}
 f(\boldsymbol{k_i}, \boldsymbol{k_j})
\equiv 2 \alpha_1 (k_i^6 + k_j^6 + k_{ij}^6) 
 + \alpha_2 (k_i^6 + k_j^6 + k_{ij}^6 - k_i^2 k_j^4
  - k_i^4 k_j^2 - k_i^2 k_{ij}^4 - k_i^4 k_{ij}^2 - k_j^2 k_{ij}^4 -
  k_j^4 k_{ij}^2) + 6 \alpha_3 k_i^2 k_j^2 k_{ij}^2 , \label{fij}
\end{equation}
with $k_{ij} \equiv \left| \boldsymbol{k_i} +  \boldsymbol{k_j}
\right|$.

\begin{figure}[htbp]
 \begin{minipage}{.42\linewidth}
  \begin{center}
\includegraphics[width=\linewidth]{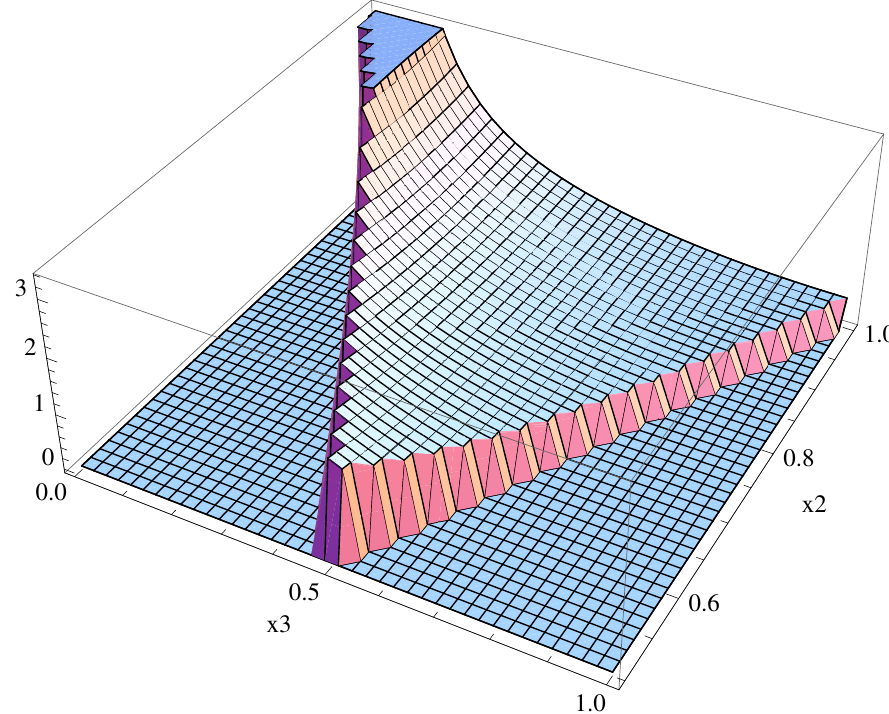}
  \end{center}
  \caption{The shape of the $\alpha_1$ contribution to $(k_1 k_2 k_3)^2 F$,
  where $\alpha_1 =-1$.} 
  \label{fig:a1}
 \end{minipage} 
 \begin{minipage}{0.05\linewidth} 
  \begin{center}
  \end{center}
 \end{minipage} 
 \begin{minipage}{.42\linewidth}
  \begin{center}
\includegraphics[width=\linewidth]{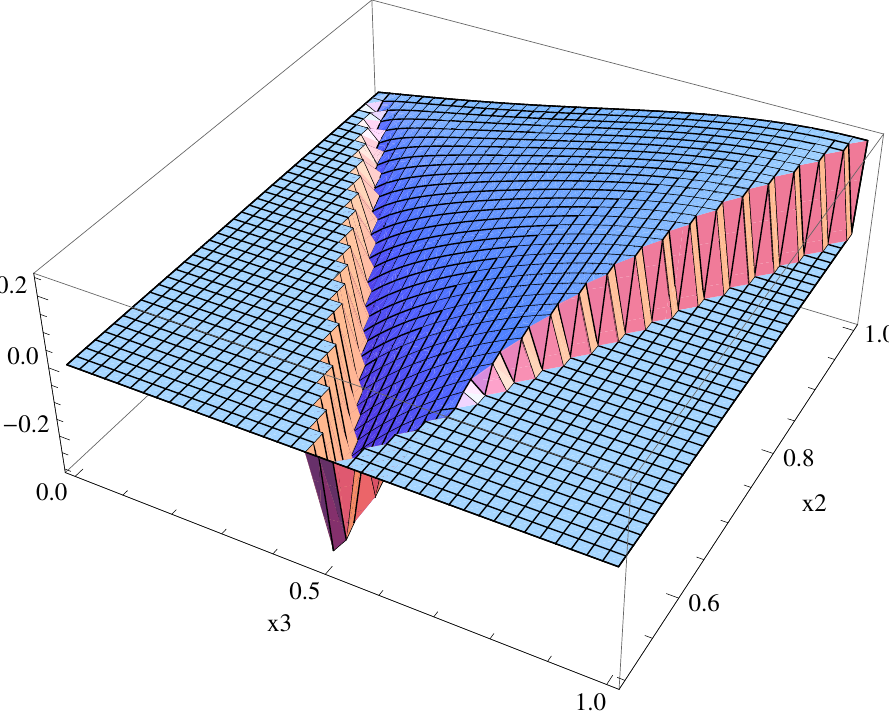}
  \end{center}
  \caption{The shape of the $\alpha_2$ contribution to $(k_1 k_2 k_3)^2 F$,
  where $\alpha_2 =1$.}
  \label{fig:a2}
 \end{minipage} 
\end{figure}

\begin{figure}[htbp]
 \begin{minipage}{.42\linewidth}
  \begin{center}
\includegraphics[width=\linewidth]{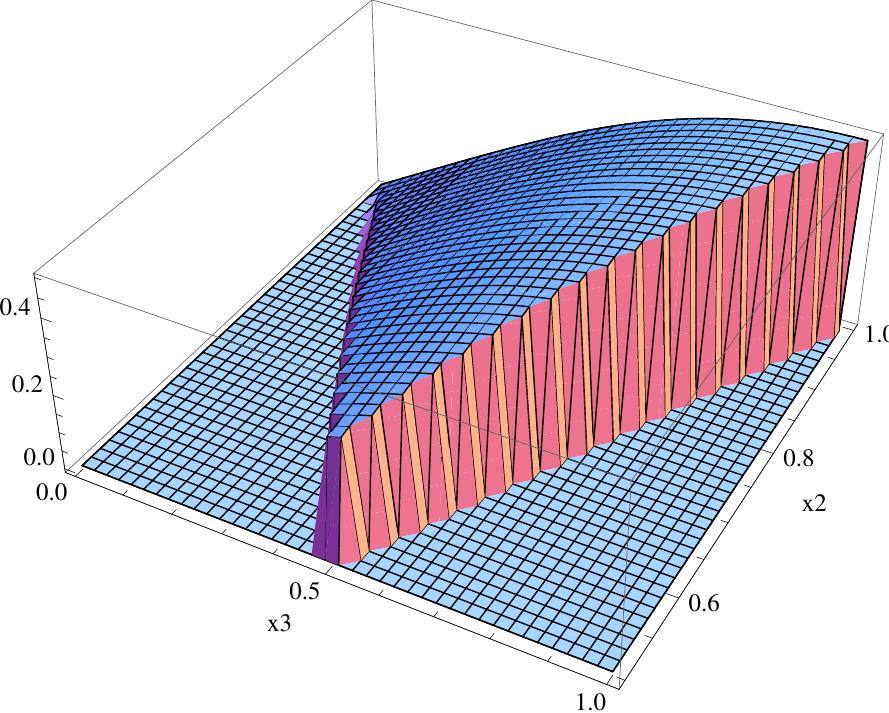}
  \end{center}
  \caption{The shape of the $\alpha_3$ contribution to $(k_1 k_2 k_3)^2 F$,
  where $\alpha_3 =-1$.}
  \label{fig:a3}
 \end{minipage} 
\end{figure}

In order to study its configuration in momentum space, let us express
the bispectrum~(\ref{bispectrum}) as
\begin{equation}
 \left\langle \phi_{\boldsymbol{k_1}} (t) \phi_{\boldsymbol{k_2}} (t)
   \phi_{\boldsymbol{k_3}} (t) \right\rangle = (2 \pi)^3 M^3 
 \delta^{(3)} (\boldsymbol{k_1} + \boldsymbol{k_2} + \boldsymbol{k_3})
 F(k_1, k_2, k_3), \label{F}
\end{equation}
and plot $ (k_1 k_2 k_3)^2 F(k_1, k_2, k_3)$ as a function of $x_2 \equiv
k_2/k_1$ and $x_3 \equiv k_3/k_1$.  
Contributions from the $\alpha_1$, $\alpha_2$, and $\alpha_3$ terms in
(\ref{fij}) are plotted respectively in Figures~\ref{fig:a1}, 
\ref{fig:a2}, and \ref{fig:a3}. 
We have assumed $0 \le x_3 \le x_2 \le 1$ to avoid showing the same
configuration twice, and $x_2 \ge 1-x_3$ further follows from the
triangular inequality.


Here, one should note that only the contribution from the $\alpha_1$
term blows up in the ``squeezed'' triangle limit, i.e. $k_3 \ll k_2
\approx k_1$, giving a shape similar to that of the local
form~\cite{Komatsu:2001rj}
\begin{equation}
 F_{\mathrm{local}}(k_1,k_2,k_3) \equiv 2 \left( \frac{1}{k_1^3 k_2^3} +
 \frac{1}{k_1^3 k_3^3} + \frac{1}{k_2^3 k_3^3} \right).
 \label{local}
\end{equation}
This can be understood as follows: Among the three interaction terms in
(\ref{S3}), only the $\alpha_1$ term breaks the shift symmetry~$\phi \to
\phi + const.$ Such a shift symmetry makes the theory indifferent to
infinitely large scale fluctuations, and suppresses interactions with
long wavelength modes. This is why $(k_1 k_2 k_3)^2 F$ from the
$\alpha_2$ and $\alpha_3$ contributions disappear in the squeezed limit. 

Moreover, the shapes from the $\alpha_2$ and $\alpha_3$ terms respectively
look similar to that of the orthogonal~\cite{Senatore:2009gt} and
equilateral~\cite{Creminelli:2005hu} 
forms which are given by
\begin{equation}
 F_{\mathrm{orthog.}} (k_1,k_2,k_3) \equiv 6 \left(
 -\frac{3}{k_1^3 k_2^3} - \frac{3}{ k_1^3 k_3^3} - \frac{3}{ k_2^3
 k_3^3} - \frac{8}{ k_1^2 k_2^2 k_3^2} + \frac{3}{ k_1 k_2^2 k_3^3} +
 (5\, \mathrm{perm.})\right),
 \label{orthog}
\end{equation}
\begin{equation}
 F_{\mathrm{equil.}} (k_1,k_2,k_3) \equiv 6 \left(
 -\frac{1}{k_1^3 k_2^3} - \frac{1}{ k_1^3 k_3^3} - \frac{1}{ k_2^3
 k_3^3} - \frac{2}{ k_1^2 k_2^2 k_3^2} + \frac{1}{ k_1 k_2^2 k_3^3} +
 (5\, \mathrm{perm.})\right).
 \label{equil}
\end{equation}
The permutations act only on the last terms in the parentheses.  

To give a bit more quantitative discussion on shapes, let us
express (\ref{bispectrum}) in terms of the above
forms. Following~\cite{Babich:2004gb,Senatore:2009gt}, we introduce a
scalar product between two distributions~$F_1$ and $F_2$,
\begin{equation}
 F_1 \cdot F_2 \equiv \sum_{\boldsymbol{k_i}} \frac{F_1(k_1,k_2,k_3)
  F_2(k_1,k_2,k_3)}{P_{k_1} P_{k_2} P_{k_3}}  \propto
 \int^1_0 dx_2 \int^1_{1-x_2} dx_3\,  x_2^4 x_3^4 F_1(1,x_2,x_3)
 F_2(1,x_2,x_3),
 \label{scpr}
\end{equation}
where summation is taken over all $\boldsymbol{k_i}$'s which form a
triangle, and $P_k \propto k^{-3}$ denotes the power spectrum. Then one
can ``expand'' $F(k_1, k_2, k_3)$ of (\ref{F}) in terms of
(\ref{local}), (\ref{orthog}), (\ref{equil}) and obtain a template function:
\begin{equation}
 F_{\mathrm{template}}(k_1, k_2, k_3) = 
 c_{\mathrm{NL}}^{\mathrm{local}} F_{\mathrm{local}}(k_1,k_2,k_3) + 
 c_{\mathrm{NL}}^{\mathrm{orthog.}} F_{\mathrm{orthog.}}(k_1,k_2,k_3) +
 c_{\mathrm{NL}}^{\mathrm{equil.}}
 F_{\mathrm{equil.}}(k_1,k_2,k_3),
 \label{template}
\end{equation}
where
\begin{align}
 & c_{\mathrm{NL}}^{\mathrm{local}} = 
 \frac{F \cdot F_{\mathrm{local}}}{F_{\mathrm{local}} \cdot
 F_{\mathrm{local}}} =
 -0.125 \alpha_1,  \label{tildefNLlocal}  \\
 & c_{\mathrm{NL}}^{\mathrm{orthog.}} = 
 \frac{F \cdot F_{\mathrm{orthog.}}}{F_{\mathrm{orthog.}} \cdot
 F_{\mathrm{orthog.}}} =
 0.226\alpha_1 + 0.0186\alpha_2 -0.00334 \alpha_3, 
 \label{tildefNLorthog}   \\
 & c_{\mathrm{NL}}^{\mathrm{equil.}} = 
 \frac{F \cdot F_{\mathrm{equil.}}}{F_{\mathrm{equil.}} \cdot
 F_{\mathrm{equil.}}} =
 -0.223\alpha_1 + 0.0280\alpha_2 -0.0876 \alpha_3.
 \label{tildefNLequil}
\end{align}
$\alpha_2$ and $\alpha_3$ are absent in (\ref{tildefNLlocal}) since the
denominator~$F_{\mathrm{local}} \cdot F_{\mathrm{local}}$ blows up.\footnote{It
is possible to avoid such divergences by introducing cutoffs in the
integration, e.g., $x_2, x_3 \ge 0.001$ which roughly corresponds to the
current observable limit in the CMB.} Here, since (\ref{local}),
(\ref{orthog}), and (\ref{equil}) do \textit{not} form a complete basis
set, one should consider (\ref{template}) as an indicator roughly
telling which combination of the 3-point interaction terms in (\ref{S3})
gives which bispectrum shape. 

Using the above template function, one can estimate the non-Gaussianity
parameter~$f_{\mathrm{NL}}$ of the primordial curvature perturbations. 
As discussed around (\ref{phiphi-to-zetazeta}),
we assume that the Lifshitz scalar fluctuations are linearly
converted to the curvature perturbations, i.e.
\begin{equation}
 \zeta = \frac{\phi}{\mu} , \label{linconv}
\end{equation}
with some mass scale~$\mu$. 
For the bispectrum of Bardeen's
curvature perturbations~$\Psi$ (which is related to the primordial
curvature perturbations by $\zeta = \frac{5}{3} \Psi$ in the
matter-dominated era)
\begin{equation}
 \langle \Psi_{\boldsymbol{k_1}} \Psi_{\boldsymbol{k_2}}
  \Psi_{\boldsymbol{k_3}} \rangle =  
 (2 \pi)^3 \delta^{(3)} (\boldsymbol{k_1} + \boldsymbol{k_2}  +
 \boldsymbol{k_3}) F_{\Psi} (k_1, k_2, k_3),
\end{equation}
we define its non-Gaussianity parameter~$f_{\mathrm{NL}}$ as
\begin{equation}
 F_{\Psi}(k,k,k) = f_{\mathrm{NL}} \Delta_{\Psi}^2 \frac{6}{ k^6}.
\end{equation}
Here, $\Delta_{\Psi}$ is the amplitude of the power spectrum
\begin{equation}
 \langle \Psi_{\boldsymbol{k_1}} \Psi_{\boldsymbol{k_2}} \rangle  = 
  (2 \pi)^3 \delta^{(3)} (\boldsymbol{k_1} + \boldsymbol{k_2} )
  \frac{\Delta_{\Psi}}{ k_1^3},
\end{equation}
which in the present case is
\begin{equation}
 \Delta_{\Psi} = \frac{9}{50} \left( \frac{M}{\mu}\right)^2 \simeq 1.7
  \times 10^{-8}.
\end{equation}
The value on the far right hand side is set by the COBE
normalization~(\ref{COBEnormalization}). Then, by substituting the template
function~(\ref{template}) into the Lifshitz scalar
fluctuations~(\ref{F}), we obtain
\begin{equation}
 f_{\mathrm{NL}}^i = \frac{20}{3} \frac{\mu}{M} c_{\mathrm{NL}}^i
  \simeq 2.2\times 10^4 c_{\mathrm{NL}}^i,
\end{equation}
where $i =$ local, orthog., equil.
We can now use this relation to set constraints on the self-coupling
strengths, by computing the necessary conditions for (\ref{tildefNLlocal}),
(\ref{tildefNLorthog}), and (\ref{tildefNLequil}) to satisfy the WMAP
7yr~\cite{Komatsu:2010fb} constraints on the non-Gaussianity parameters:
$-10 < f_{\mathrm{NL}}^{\mathrm{local}} < 74$, 
$ -410 < f_{\mathrm{NL}}^{\mathrm{orthog.}} < 6$, and 
$ -214 < f_{\mathrm{NL}}^{\mathrm{equil.}} < 266$ (each at 95\% CL),
respectively. Thus we arrive at
\begin{equation}
 -0.03 < \alpha_1 < 0.004, \qquad
 -1 < \alpha_2 < 0.4, \qquad
 -0.5 < \alpha_3 < 0.3.
 \label{alphabound}
\end{equation}
We expect that more rigorous treatments (such as using observational
data to constrain combinations of different shapes, instead of
individual ones) would not change the results significantly.\footnote{We
also note that (\ref{alphabound}) are conservative bounds, since they
are necessary conditions for satisfying observational constraints. 
One way to obtain more strict bounds is by imposing observational
constraints under an \textit{ad hoc} assumption
that two out of the three $\alpha_i$'s are zero. However,
such procedure does not change the results significantly, except for the
strong upper limit on $f_{\mathrm{NL}}^{\mathrm{orthog.}}$ tightening
the upper bound on $\alpha_2$ and lower bound on $\alpha_3$ roughly by
an order of magnitude.} 

The bounds in (\ref{alphabound}) show that observational constraints
require $|\alpha_2 |, |\alpha_3 | = O(10^{-1})$. 
Note that such smallness of the coupling strengths is required anyway
for validity of the perturbative expansions in terms of~$H_I$, which we
have carried out in computing the bispectrum. In other words, as $\alpha_2$ or
$\alpha_3$ become of order unity and our procedure for computing the bispectrum
breaks down, the curvature perturbations saturate the
current observational limit for the orthogonal/equilateral form
bispectra. The most stringent bounds are obtained for $\alpha_1$,
i.e. $|\alpha_1 | = O(10^{-2} \mathrm{-} 10^{-3})$ where the required
level of tuning depends on its sign. This is mainly due to the rather
tight constraints on the local form bispectrum it produces.
In order to suppress $\alpha_1$, some sort of symmetry may be required
in the theory. For example, as we have stated below~(\ref{local}), a
shift symmetry for $\phi$ forbids such self-coupling terms producing
local-type bispectra.  

Here we have focused on non-Gaussianity sourced by the intrinsic
fluctuations of the Lifshitz scalar, but we should remark that further
non-Gaussianities in the resulting curvature perturbations can be
generated through non-linear conversion processes.

\subsection{Trispectrum}
\label{subsec:trispectrum}

\begin{figure}[htbp]
 \begin{minipage}{.35\linewidth}
  \begin{center}
\includegraphics[width=\linewidth]{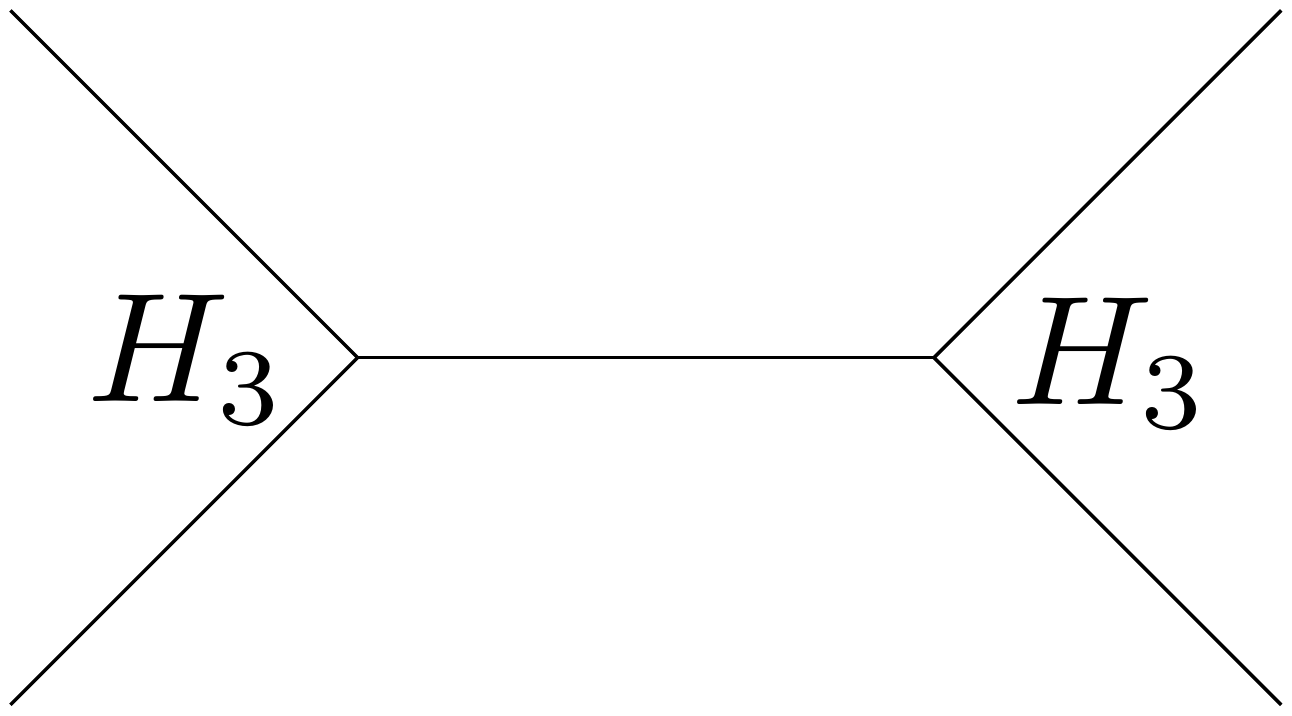}
  \end{center}
  \caption{The scalar-exchange diagram.}
  \label{fig:se}
 \end{minipage} 
 \begin{minipage}{0.05\linewidth} 
  \begin{center}
  \end{center}
 \end{minipage} 
 \begin{minipage}{.35\linewidth}
  \begin{center}
\includegraphics[width=.56\linewidth]{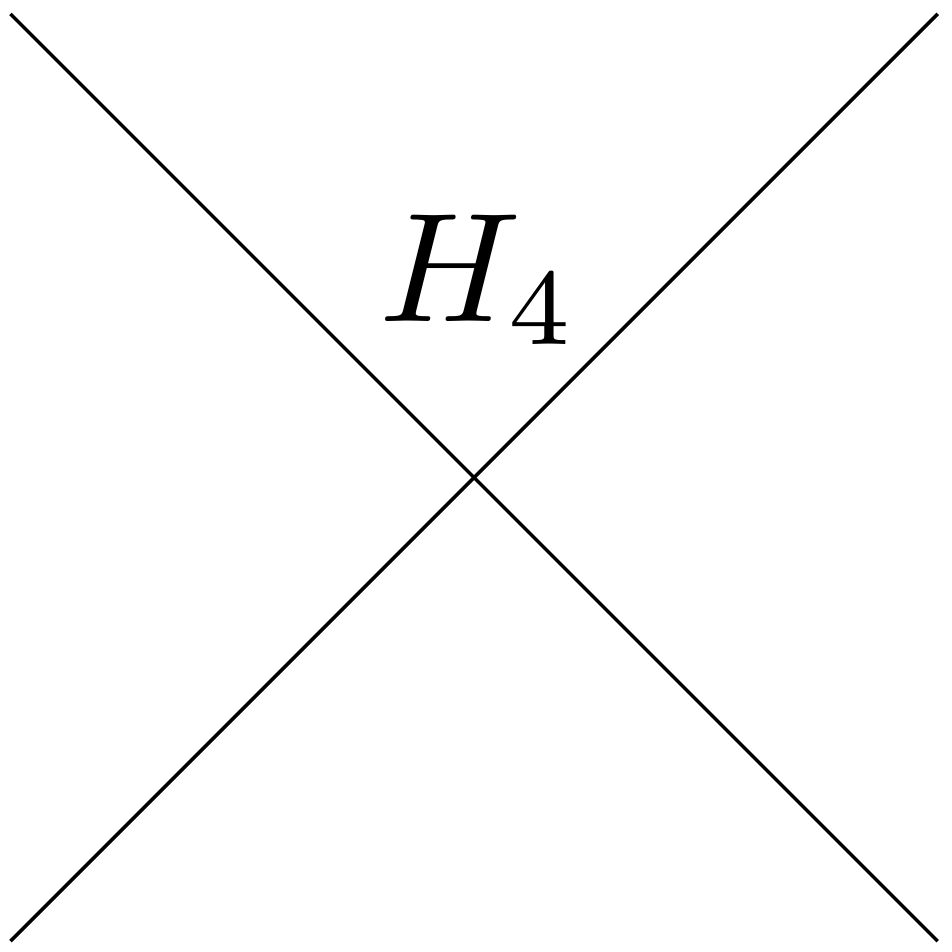}
  \end{center}
  \caption{The contact-interaction diagram.}
  \label{fig:ci}
 \end{minipage} 
\end{figure}

The trispectrum can be obtained in a similar manner to the previous
subsection. We compute contributions from the scalar-exchange diagram
(Figure~\ref{fig:se}), and from the contact-interaction diagram
(Figure~\ref{fig:ci}).
The 4-point interaction terms of the Lifshitz scalar action are given in 
Appendix~\ref{appendix} as 
\begin{eqnarray}
&& S_4 = \int dt d^3x \, \frac{1}{M^6 a(t)^3} 
 \bigl\{ \beta_1 \phi^3 \Delta^3 \phi + \beta_2 \phi^2 (\Delta \phi)
  (\Delta^2 \phi) + \beta_3 \phi ( \Delta \phi)^3 \nonumber\\
&&\qquad\qquad\qquad\qquad\qquad\qquad\qquad
  + \beta_4 \phi^2
  (\Delta \partial_i \phi)^2 
  + \beta_5 \phi^2 (\partial_i \partial_j
  \partial_k \phi)^2 + \beta_6 (\partial_i \partial_j \partial_k \phi)
  (\partial_i \phi) (\partial_j \phi) (\partial_k \phi)\bigr\},
 \label{S4}
\end{eqnarray} 
where $\beta_i$ are dimensionless parameters.
Therefore the 4-point interaction Hamiltonian becomes
\begin{eqnarray}
&&  H_4(t) = -\int  d^3x \, \frac{1}{M^6 a(t)^3} 
 \bigl\{ \beta_1 \phi^3 \Delta^3 \phi + \beta_2 \phi^2 (\Delta \phi)
  (\Delta^2 \phi) + \beta_3 \phi ( \Delta \phi)^3 \nonumber\\
&&\qquad\qquad\qquad\qquad\qquad\qquad\qquad
  + \beta_4 \phi^2
  (\Delta \partial_i \phi)^2 
  + \beta_5 \phi^2 (\partial_i \partial_j
  \partial_k \phi)^2 + \beta_6 (\partial_i \partial_j \partial_k \phi)
  (\partial_i \phi) (\partial_j \phi) (\partial_k \phi)\bigr\}.
\end{eqnarray}
Even if the 3-point interactions are suppressed, i.e. $|\alpha_i | \ll
1$, the 4-point interaction terms can produce a large trispectrum
through the contact-interaction diagram. 

The contribution from the scalar-exchange diagram are obtained as
follows (Note that we only compute contributions from connected diagrams.),
\begin{align}
 \left\langle \phi_{\boldsymbol{k_1}} (t) \phi_{\boldsymbol{k_2}} (t)
   \phi_{\boldsymbol{k_3}} (t) \phi_{\boldsymbol{k_4}} (t)
 \right\rangle_{\mathrm{s.e.}} 
 & = \left. \left(
 - \int^{t}_{t_0} dt' \int^{t'}_{t_0} d t'' \langle \left[ H_{3} (t''),
 \left[ H_3(t') ,
 \phi_{\boldsymbol{k_1}} (t) \phi_{\boldsymbol{k_2}} (t)
   \phi_{\boldsymbol{k_3}} (t)  \phi_{\boldsymbol{k_4}} (t)  \right]
 \right] \rangle \right) \right|_{\mathrm{connected}}
 \\ 
 & =  \frac{(2 \pi )^3 M^4}{2^3} 
 \frac{\delta^{(3)} (\boldsymbol{k_1} + \boldsymbol{k_2} +
 \boldsymbol{k_3} +  \boldsymbol{k_4} )
 (k_1^3 + k_2^3 + k_3^3 + k_4^3 + k_{12}^3) 
 f(\boldsymbol{k_1}, \boldsymbol{k_2}) f(\boldsymbol{k_3},
 \boldsymbol{k_4})}{
 (k_1 k_2 k_3 k_4 k_{12})^3 (k_1^3 + k_2^3+ k_{12}^3) (k_3^3 + k_4^3 +
 k_{12}^3) (k_1^3 + k_2^3 + k_3^3 + k_4^3) } \\
 & \qquad  + \biggl(
 \mathrm{2\, \, terms\, \, with\, \, }
 (1,2)(3,4) \longrightarrow (1,3)(2,4),\,  (1,4)(2,3)
\biggr), \label{tri-se}
\end{align}
where $ f(\boldsymbol{k_i}, \boldsymbol{k_j}) $ is defined in
(\ref{fij}). One can also compute the contribution from the
contact-interaction diagram,
\begin{align}
 \left\langle \phi_{\boldsymbol{k_1}} (t) \phi_{\boldsymbol{k_2}} (t)
   \phi_{\boldsymbol{k_3}} (t) \phi_{\boldsymbol{k_4}} (t)
 \right\rangle_{\mathrm{c.i.}} 
 & = \left. \left( i \int^{t}_{t_0} dt' \langle \left[ H_{4} (t'),
 \phi_{\boldsymbol{k_1}} (t) \phi_{\boldsymbol{k_2}} (t)
   \phi_{\boldsymbol{k_3}} (t)  \phi_{\boldsymbol{k_4}} (t) \right]
 \rangle \right) \right|_{\mathrm{connected}} \\ 
 & = - \frac{(2\pi)^3 M^4}{2^3} \frac{ \delta^{(3)} (\boldsymbol{k_1} +
 \boldsymbol{k_2} + \boldsymbol{k_3} + \boldsymbol{k_4} )
 r(\boldsymbol{k_1}, \boldsymbol{k_2}, \boldsymbol{k_3},
 \boldsymbol{k_4})  }{(k_1 k_2 k_3 k_4)^3 (k_1^3 + k_2^3 +  k_3^3 +
 k_4^3)},
 \label{tri-ci}
\end{align}
where we have defined 
\begin{equation}
\begin{split}
 &
 r(\boldsymbol{k_1},\boldsymbol{k_2},\boldsymbol{k_3},\boldsymbol{k_4}) 
 \equiv \\
 & \quad \, \,   6 \beta_1 \left(k_1^6 + k_2^6 + k_3^6 + k_4^6 \right) \\
 & + 2 \beta_2 \left(
 k_1^2 k_2^4  + k_1^4 k_2^2 + k_1^2 k_3^4 + k_1^4 k_3^2 + k_1^2 k_4^4 +
 k_1^4 k_4^2 + k_2^2 k_3^4 + k_2^4 k_3^2 + k_2^2 k_4^4 + k_2^4 k_4^2 +
 k_3^2 k_4^4 + k_3^4 k_4^2 \right) \\
 & + 6 \beta_3 \left( k_1^2 k_2^2 k_3^3 + k_1^2 k_2^2 k_4^2 + k_1^2 k_3^2
	      k_4^2 + k_2^2 k_3^2 k_4^2
 \right) \\
 & + 4 \beta_4 \left( k_1^2 k_2^2 (\boldsymbol{k_1}\cdot \boldsymbol{k_2}) +
	      k_1^2 k_3^2 (\boldsymbol{k_1}\cdot \boldsymbol{k_3}) + k_1^2
	      k_4^2 (\boldsymbol{k_1}\cdot \boldsymbol{k_4}) + k_2^2 k_3^2
	      (\boldsymbol{k_2}\cdot \boldsymbol{k_3}) + k_2^2 k_4^2
	      (\boldsymbol{k_2}\cdot \boldsymbol{k_4}) + k_3^2 k_4^2
	      (\boldsymbol{k_3}\cdot \boldsymbol{k_4})
 \right) \\
 & + 4 \beta_5  \left(
  (\boldsymbol{k_1} \cdot \boldsymbol{k_2})^3
 + (\boldsymbol{k_1} \cdot \boldsymbol{k_3})^3
 + (\boldsymbol{k_1} \cdot \boldsymbol{k_4})^3
 + (\boldsymbol{k_2} \cdot \boldsymbol{k_3})^3
 + (\boldsymbol{k_2} \cdot \boldsymbol{k_4})^3
 + (\boldsymbol{k_3} \cdot \boldsymbol{k_4})^3
\right) \\
 & + 6 \beta_6 \, (
 (\boldsymbol{k_1} \cdot \boldsymbol{k_2})(\boldsymbol{k_1} \cdot
 \boldsymbol{k_3})(\boldsymbol{k_1} \cdot \boldsymbol{k_4}) 
 +  (\boldsymbol{k_2} \cdot \boldsymbol{k_1})(\boldsymbol{k}_2 \cdot
 \boldsymbol{k_3})(\boldsymbol{k_2} \cdot \boldsymbol{k_4}) \\
 & \qquad \qquad 
 +  (\boldsymbol{k_3} \cdot \boldsymbol{k_1})(\boldsymbol{k_3} \cdot 
 \boldsymbol{k_2})(\boldsymbol{k_3} \cdot \boldsymbol{k_4}) 
 +  (\boldsymbol{k_4} \cdot \boldsymbol{k_1})(\boldsymbol{k_4} \cdot
 \boldsymbol{k_2})(\boldsymbol{k_4} \cdot \boldsymbol{k_3}) 
 ).
\end{split}
\end{equation}

\begin{figure}[htbp]
 \begin{minipage}{.42\linewidth}
  \begin{center}
\includegraphics[width=\linewidth]{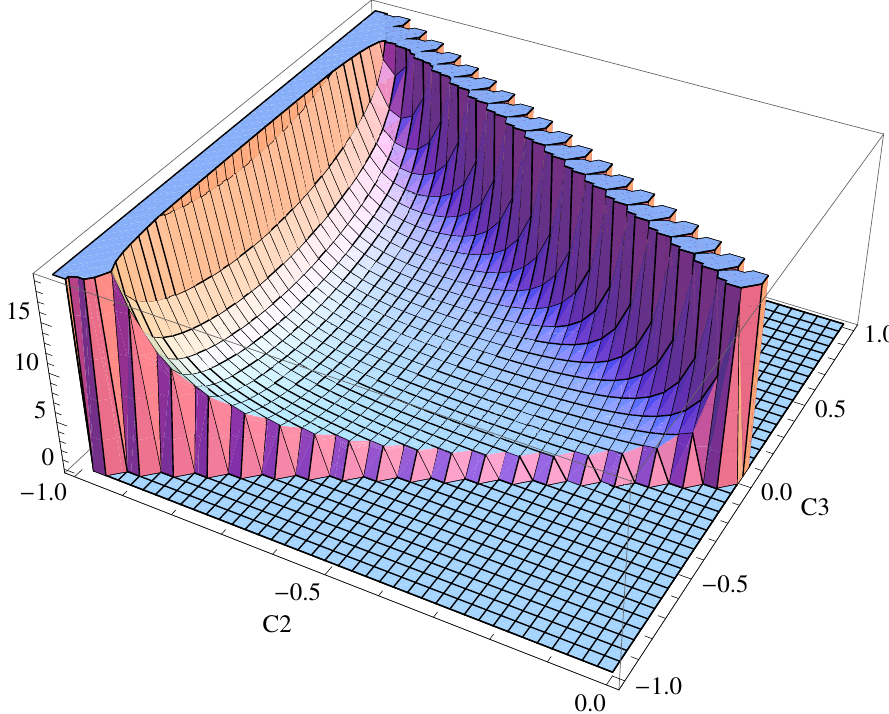}
  \end{center}
  \caption{The shape of the $\alpha_1^2$ contribution to $(k_1 k_2 k_3
  k_4)^{9/4} \mathcal{T}_{\mathrm{s.e.}}$, where $\alpha_1^2 =1$.} 
  \label{fig:a1a1}
 \end{minipage} 
 \begin{minipage}{0.05\linewidth} 
  \begin{center}
  \end{center}
 \end{minipage} 
 \begin{minipage}{.42\linewidth}
  \begin{center}
\includegraphics[width=\linewidth]{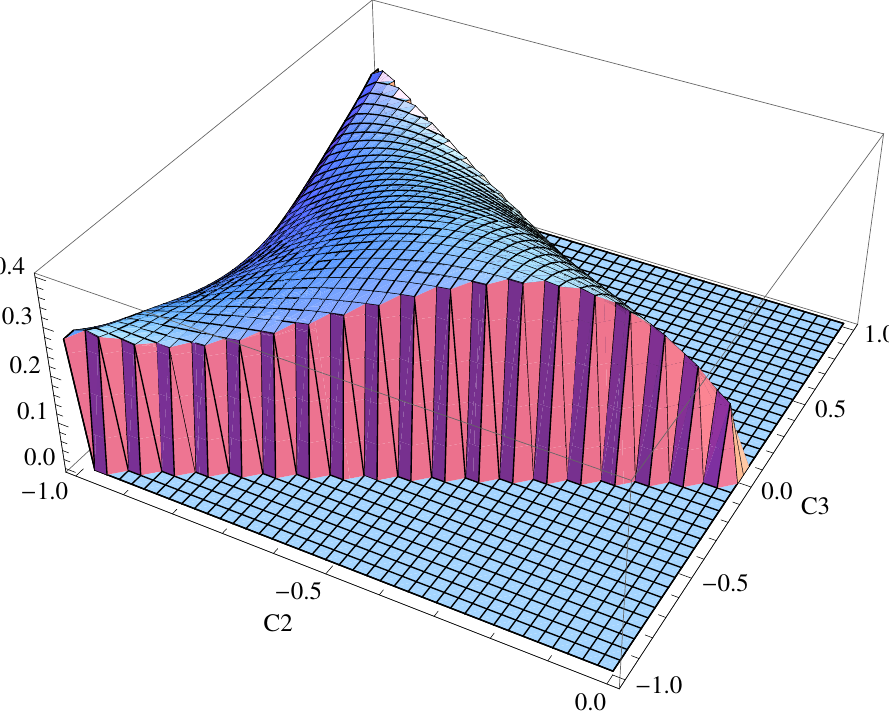}
  \end{center}
  \caption{The shape of the $\alpha_2^2$ contribution to $(k_1 k_2 k_3
  k_4)^{9/4} \mathcal{T}_{\mathrm{s.e.}}$, where $\alpha_2^2 =1$.}
  \label{fig:a2a2}
 \end{minipage} 
\end{figure}

\begin{figure}[htbp]
 \begin{minipage}{.42\linewidth}
  \begin{center}
\includegraphics[width=\linewidth]{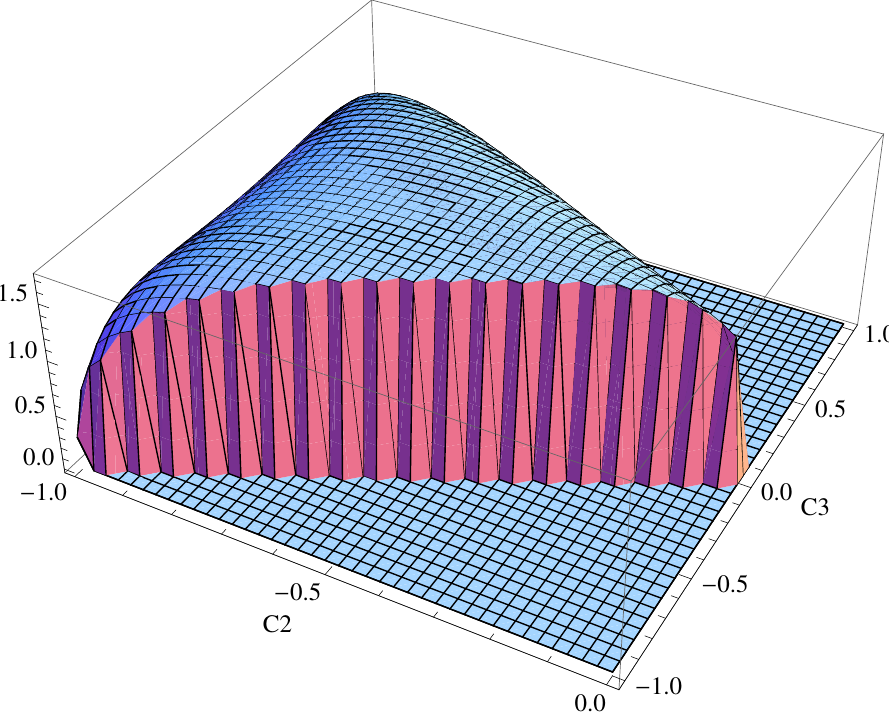}
  \end{center}
  \caption{The shape of the $\alpha_3^2$ contribution to $(k_1 k_2 k_3
  k_4)^{9/4} \mathcal{T}_{\mathrm{s.e.}}$, where $\alpha_3^2 =1$.}
  \label{fig:a3a3}
 \end{minipage} 
 \begin{minipage}{0.05\linewidth} 
  \begin{center}
  \end{center}
 \end{minipage} 
 \begin{minipage}{.42\linewidth}
  \begin{center}
\includegraphics[width=\linewidth]{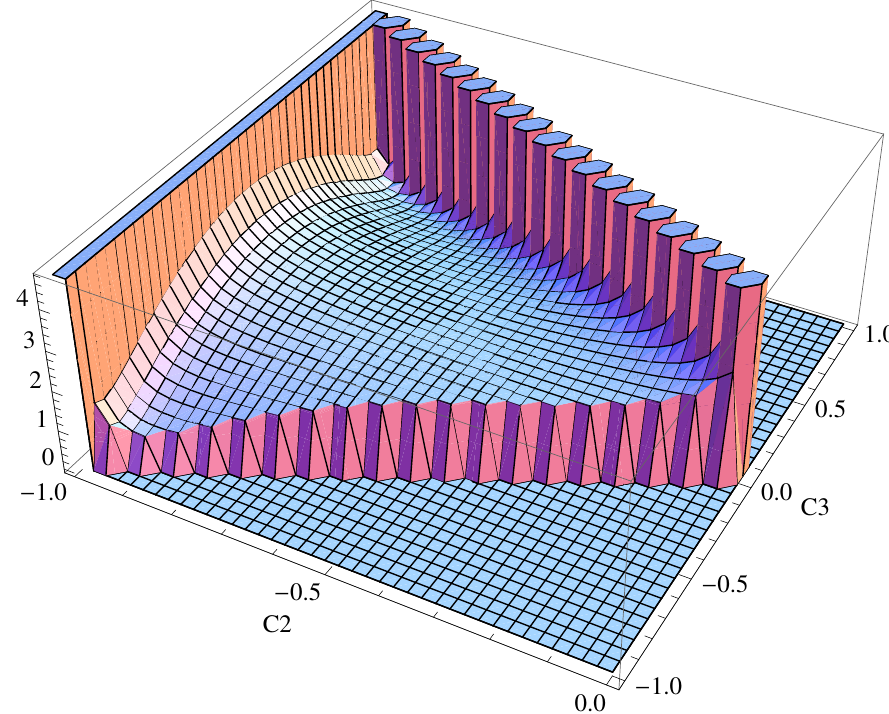}
  \end{center}
  \caption{The shape of the $\alpha_1 \alpha_2$ contribution to $(k_1
  k_2 k_3  k_4)^{9/4} \mathcal{T}_{\mathrm{s.e.}}$, where $\alpha_1 \alpha_2 =- 1$.}
  \label{fig:a1a2}
 \end{minipage} 
\end{figure}

\begin{figure}[htbp]
 \begin{minipage}{.42\linewidth}
  \begin{center}
\includegraphics[width=\linewidth]{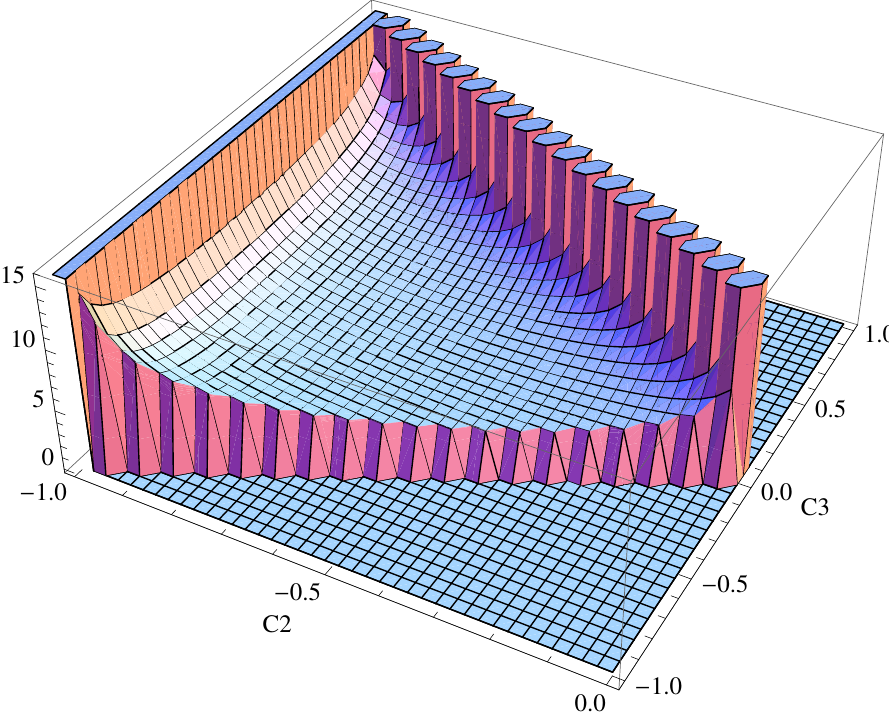}
  \end{center}
  \caption{The shape of the $\alpha_1 \alpha_3$ contribution to $(k_1
  k_2 k_3  k_4)^{9/4} \mathcal{T}_{\mathrm{s.e.}}$, where $\alpha_1 \alpha_3 = 1$.}
  \label{fig:a1a3}
 \end{minipage} 
 \begin{minipage}{0.05\linewidth} 
  \begin{center}
  \end{center}
 \end{minipage} 
 \begin{minipage}{.42\linewidth}
  \begin{center}
\includegraphics[width=\linewidth]{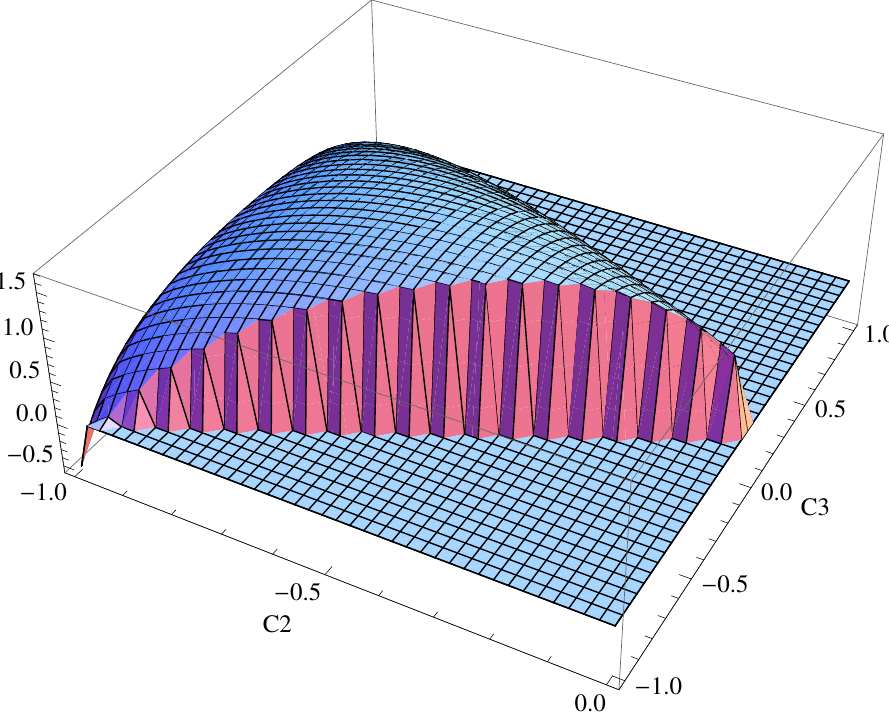}
  \end{center}
  \caption{The shape of the $\alpha_2 \alpha_3$ contribution to $(k_1
  k_2 k_3  k_4)^{9/4} \mathcal{T}_{\mathrm{s.e.}}$, where $\alpha_2 \alpha_3 = -1$.}
  \label{fig:a2a3}
 \end{minipage} 
\end{figure}

\begin{figure}[htbp]
 \begin{minipage}{.42\linewidth}
  \begin{center}
\includegraphics[width=\linewidth]{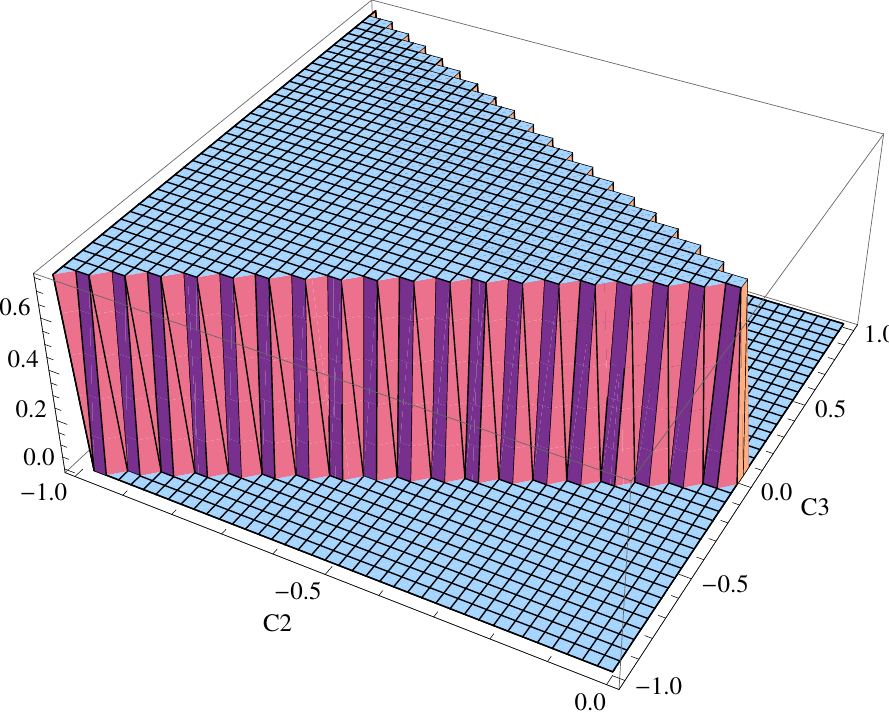}
  \end{center}
  \caption{The shape of the $\beta_1$ contribution to $(k_1
  k_2 k_3  k_4)^{9/4} \mathcal{T}_{\mathrm{c.i.}}$, where $\beta_1 = -1$.}
  \label{fig:b1}
 \end{minipage} 
 \begin{minipage}{0.05\linewidth} 
  \begin{center}
  \end{center}
 \end{minipage} 
 \begin{minipage}{.42\linewidth}
  \begin{center}
\includegraphics[width=\linewidth]{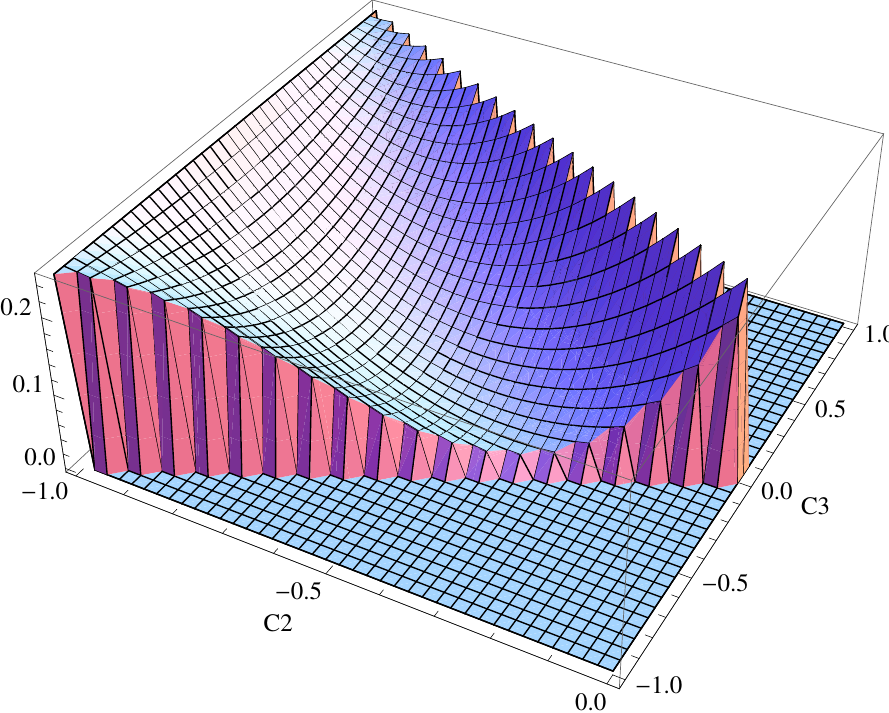}
  \end{center}
  \caption{The shape of the $\beta_5$ contribution to $(k_1
  k_2 k_3  k_4)^{9/4} \mathcal{T}_{\mathrm{c.i.}}$, where $\beta_5 = 1$.}
  \label{fig:b5}
 \end{minipage} 
\end{figure}

\begin{figure}[htbp]
 \begin{minipage}{.42\linewidth}
  \begin{center}
\includegraphics[width=\linewidth]{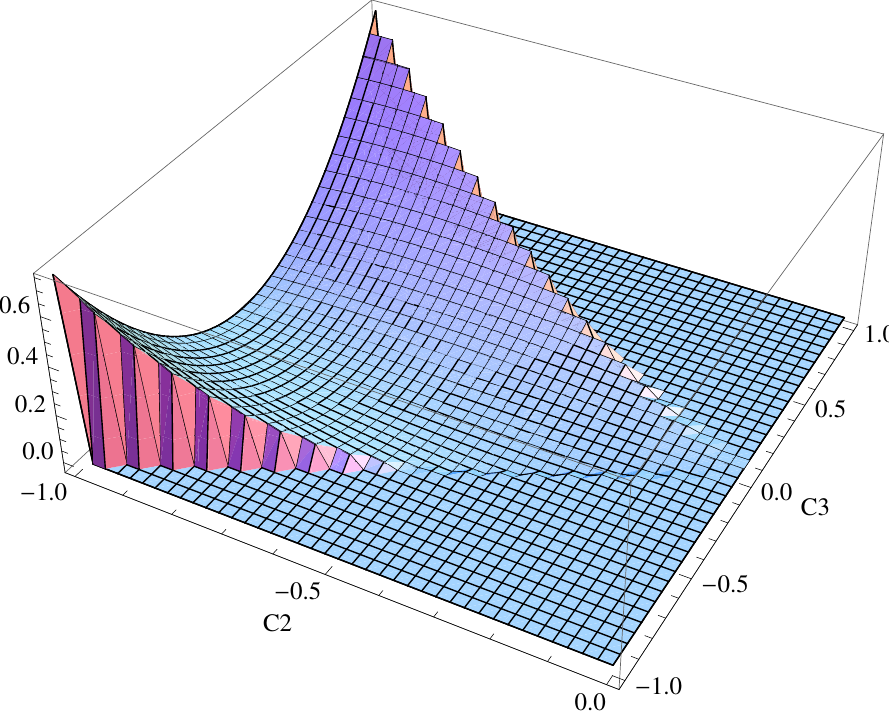}
  \end{center}
  \caption{The shape of the $\beta_6$ contribution to $(k_1
  k_2 k_3  k_4)^{9/4} \mathcal{T}_{\mathrm{c.i.}}$, where $\beta_6 = -1$.}
  \label{fig:b6}
 \end{minipage} 
\end{figure}

Expressing the trispectra as 
\begin{equation}
 \left\langle \phi_{\boldsymbol{k_1}} (t) \phi_{\boldsymbol{k_2}} (t)
   \phi_{\boldsymbol{k_3}} (t) \phi_{\boldsymbol{k_4}} (t) \right\rangle
 = (2 \pi)^3 M^4  \delta^{(4)} (\boldsymbol{k_1} + \boldsymbol{k_2} +
 \boldsymbol{k_3} + \boldsymbol{k_4}) 
 \mathcal{T}(\boldsymbol{k_1}, \boldsymbol{k_2}, \boldsymbol{k_3},
 \boldsymbol{k_4}), \label{T} 
\end{equation}
let us plot $(k_1 k_2 k_3 k_4)^{9/4} \mathcal{T}(\boldsymbol{k_1},
\boldsymbol{k_2}, \boldsymbol{k_3},  \boldsymbol{k_4})$ in the
``equilateral'' limit:~$k_1 = k_2 = k_3 = k_4$. In this limit, the shape
of the tetrahedron formed by $\boldsymbol{k_i}$'s depends on two
independent variables, e.g., $C_2 \equiv \boldsymbol{k_1} \cdot
\boldsymbol{k_2}/{k_1 k_2}$ and $C_3 \equiv \boldsymbol{k_1} \cdot
\boldsymbol{k_3}/{k_1 k_3}$. Contributions to
$\mathcal{T}_{\mathrm{s.e.}}$~(\ref{tri-se}) from the $\alpha_1^2$,
$\alpha_2^2$, $\alpha_3^2$, $\alpha_1 \alpha_2$, $\alpha_1 \alpha_3$,
and $\alpha_2 \alpha_3$ terms are plotted respectively in
Figures~\ref{fig:a1a1}, \ref{fig:a2a2}, \ref{fig:a3a3}, \ref{fig:a1a2},
\ref{fig:a1a3}, and \ref{fig:a2a3}. Contributions to
$\mathcal{T}_{\mathrm{c.i.}}$~(\ref{tri-ci}) from the $\beta_1$,
$\beta_5$, $\beta_6$ terms are plotted in Figures~\ref{fig:b1},
\ref{fig:b5}, and \ref{fig:b6}. We have omitted contributions from the
$\beta_2$, $\beta_3$, $\beta_4$ terms since in the equilateral limit
they become flat and are equivalent to that from $\beta_1$ up to overall
constant factors. The condition $C_2 + C_3 \le 0$ is required for
$\boldsymbol{k_i}$'s to close, and we further assume $C_2 \le C_3$ in
order to avoid showing the same configuration twice. 

\clearpage


\clearpage

\section{Conclusions}
\label{sec:summary}

In this paper, we have studied non-Gaussianity in the intrinsic
fluctuations of a Lifshitz scalar which follows an anisotropic scaling
with $z = 3$. Our work is based on~\cite{Mukohyama:2009gg}, which
pointed out that its special dispersion relation in the UV can lead to
generation of super-horizon field perturbations. 
Since the scaling dimension of a Lifshitz scalar with $z=3$ is zero,
the resulting field perturbations become scale-invariant whether or not
the scalar's self-couplings are small. This leads to our main point that
curvature perturbations generated from such field fluctuations
necessarily leave large non-Gaussianity in the sky, 
unless the field's self-couplings are forbidden by some symmetry, or the
field exhibits some sort of asymptotic freedom. This is to be contrasted
with perturbations generated through cosmic inflation, where largely
non-Gaussian intrinsic fluctuations are in most cases incompatible with
scale-invariance. 

The Lifshitz scalar's self-coupling terms containing spatial derivatives
produce non-Gaussianities with various configurations in momentum
space. In particular, the bispectrum of the field fluctuations includes
shapes which are similar to that of the local, equilateral, and orthogonal
forms. (However, we emphasize that the local, equilateral, and
orthogonal shapes do not form a complete basis set for the bispectrum
obtained in this paper. We also note that the results of the effective
field theory approach in \cite{Cheung:2007st} do not apply to our case,
where Lorentz symmetry is explicitly broken, and non-Gaussianity is
sourced by marginal terms in the action.) Upon computing the correlation
functions, we have carried out expansions in terms of the interaction
Hamiltonian. Within the domain of applicability of such perturbative
expansion, i.e. the self-couplings less than unity, we have seen that
the Lifshitz scalar's field fluctuations can lead to significant
non-Gaussianity in the primordial curvature perturbations. In
particular, when curvature perturbations are sourced linearly from the
field fluctuations as in (\ref{linconv}), their bispectrum saturates the
current observational limit for the orthogonal and equilateral forms, as
the self-couplings $\alpha_2$ and $\alpha_3$ in (\ref{S3}) approach
unity. Since naively there is no reason for such self-couplings to be
suppressed, we can expect large non-Gaussianity to be produced from
Lifshitz scalar fluctuations, which may be detected by upcoming CMB
observations. On the other hand, for the local-type bispectrum,
observational constraints require $\alpha_1$ to be as small as
$O(10^{-2} \mathrm{-} 10^{-3})$ (the level of tuning depends on
$\alpha_1$'s sign). However, as we have remarked, such self-couplings
sourcing local-type non-Gaussianity can be forbidden by a shift
symmetry. 

The field fluctuations generated in the mechanism of
\cite{Mukohyama:2009gg} obtain a scale-invariant spectrum. However,
when one takes into account the renormalization-group flow of the
parameters of the theory (e.g. $M$ in (\ref{S2})), the spectrum may
become tilted. A time-dependent background value~$\Phi_0(t)$ may also give
rise to similar effects. How strong the tilt becomes, as well as the
scale-dependence of the non-Gaussianity, remains to be
understood. While in this paper we have studied fluctuations of scalar
fields, the scalar graviton which can show up in Ho\v{r}ava-Lifshitz gravity
may also obtain fluctuations in a similar manner. It would be
interesting to investigate the possibility that such scalar graviton
generates the primordial curvature
perturbations. (Ref.~\cite{Chen:2009jr} works in this direction. See
\cite{Mukohyama:2010xz} for some issues related to the scalar graviton,
including non-perturbative continuity of the limit in which general
relativity is supposed to be recovered. See also \cite{Horava:2010zj}
for a recent attempt to eliminate the scalar graviton from the theory.) 
Furthermore, when considering cosmic inflation in Ho\v{r}ava-Lifshitz
gravity, due to the field fluctuations freezing-out at the time of sound
horizon~$(M^2 H)^{-1/3}$ exit, the well-known relations in slow-roll
inflation between various cosmological observables and the slow-roll
parameters are expected to be modified. Aspects of cosmic inflation in
Ho\v{r}ava-Lifshitz gravity are also worthy of study in details.

\begin{acknowledgments}
 We would like to thank Eiichiro Komatsu and Takahiro Nishimichi for 
 helpful conversations. T.K. is also grateful to Masahiro Kawasaki for his
 continuous support. The work of T.K. was supported by Grant-in-Aid for
 JSPS Fellows No.~21$\cdot$8966. The work of S.M. was supported in part by
 Grant-in-Aid for Young Scientists (B) No. 17740134, Grant-in-Aid for
 Creative Scientific Research No. 19GS0219, Grant-in-Aid for Scientific
 Research on Innovative Areas No. 21111006, Grant-in-Aid for Scientific
 Research (C) No. 21540278, and the Mitsubishi Foundation. This work was
 supported by World Premier International Research Center Initiative
 (WPI Initiative), MEXT, Japan. 
\end{acknowledgments}

\appendix

\section{Nonlinear terms in the UV action}
\label{appendix}

In this appendix we construct nonlinear, marginal terms in the action 
in the UV, specializing to the $z=3$ case. As mentioned after
(\ref{actionPhi}), those nonlinear terms do not include time derivatives
but can include spatial derivatives. We demand that each term in the
action can include only up to six spatial derivatives. This treatment is
justified since with $z=3$, terms with more than six spatial derivatives
would be non-renormalizable and thus would not be generated by quantum 
corrections. In this case the most important terms in the UV are
marginal ones, i.e. those with six spatial derivatives. In the following
we write down all independent combinations of cubic and quartic terms
which are marginal in the UV~\footnote{
In \cite{Gao:2009bx}, non-Gaussianity of cosmological perturbations
in Ho\v{r}ava-Lifshitz gravity is discussed. However, in that paper,
non-renormalizable interaction terms are considered as the main source
of non-Gaussianity. In the present paper, on the other hand, all terms
in the action are power-counting renormalizable and we consider marginal
ones as the main source of non-Gaussianity.}.

\subsection{Cubic terms}

In general we can write down fourteen cubic terms with six spatial
derivatives as follows,
\begin{eqnarray}
  \begin{array}{l}
       \\
  \begin{array}{llll}
A_1 \equiv \left(\Delta^3 \phi\right) \phi^2 , 
&A_2 \equiv\left(\Delta^2 \partial_i \phi\right)\left( \partial_i \phi\right) \phi ,
&A_3 \equiv\left(\Delta^2\phi\right)\left( \Delta\phi\right),
&A_4 \equiv\left(\Delta \partial_i\partial_j\phi\right)\left( \partial_i\partial_j\phi\right)
 \phi ,\\
A_5 \equiv\left(\Delta\partial_i\phi\right)\left( \Delta\partial_i\phi\right) \phi,\ 
&A_6 \equiv\left(\partial_i\partial_j\partial_k\phi\right)
\left( \partial_i\partial_j\partial_k\phi\right) \phi,\ 
&A_7 \equiv\left(\Delta^2 \phi\right)\left( \partial_i \phi\right)
\left( \partial_i \phi\right),\ 
&A_8 \equiv\left(\Delta \partial_i\partial_j \phi\right)
\left( \partial_i\phi\right)\left( \partial_j\phi\right),\  \\
  \end{array}
       \\
  \begin{array}{lll}
A_9 \equiv\left(\Delta\partial_i\phi\right)\left( \Delta\phi\right)
\left( \partial_i\phi\right) , \qquad
&A_{10} \equiv\left(\Delta\partial_i\phi\right)\left( \partial_i\partial_j\phi\right)
\left( \partial_j\phi\right) , \qquad
&A_{11} \equiv\left(\partial_i\partial_j\partial_k\phi\right) 
\left(\partial_i\partial_j\phi\right)\left( \partial_k\phi\right) , \\
A_{12} \equiv\left(\Delta\phi\right)\left( \Delta\phi\right)\left( \Delta\phi\right) 
,\qquad
&A_{13} \equiv\left(\Delta\phi\right)\left( \partial_i\partial_j\phi\right)
\left( \partial_i\partial_j\phi\right),\qquad
&A_{14} \equiv\left(\partial_i\partial_j\phi\right)\left( \partial_j\partial_k\phi\right)
\left( \partial_k\partial_i\phi\right) . \qquad\\
  \end{array}
  \end{array}
\label{A}
\end{eqnarray}
Some linear combinations of these terms turn out to be total
derivatives. Thus, not all of these fourteen terms lead to independent
terms in the action, provided that surface terms are
dropped. Apparently, there are thirteen total derivatives with six
spatial derivatives as follows, 
\begin{eqnarray}
  \begin{array}{ll}
\partial_i \left[\left(\Delta^2 \partial_i \phi\right) 
\phi^2 \right]=A_1+2A_2,
 & \partial_i \left[\left(\Delta^2 \phi\right)
\left( \partial_i\phi\right) \phi \right]=A_2+A_3+A_7,
 \\ \partial_i \left[\left(\Delta\partial_i \partial_j \phi\right)
\left(\partial_j \phi \right) \phi \right]=A_2+A_4+A_8,
 & \partial_i \left[ \left(\Delta\partial_i \phi\right) 
\left(\Delta \phi\right) \phi \right]=A_3+A_5+A_9, 
 \\ \partial_i \left[\left(\Delta\partial_j \phi\right)
\left(\partial_i\partial_j \phi\right) \phi \right]=A_4+A_5+A_{10},
 & \partial_i \left[\left(\partial_i\partial_j\partial_k \phi\right) 
\left(\partial_j \partial_k\phi\right) \phi \right]=A_4+A_6+A_{11},
 \\ \partial_i \left[\left(\Delta \partial_i \phi\right)
\left( \partial_j\phi\right)\left(\partial_j \phi\right) \right]=A_7+2A_{10},
 & \partial_i \left[\left(\Delta\partial_j \phi\right)
\left( \partial_i\phi \right)\left(\partial_j\phi\right) \right]=A_8+A_{9}+A_{10},
 \\ \partial_i \left[\left(\partial_i\partial_j\partial_k \phi\right)
\left( \partial_j \phi\right)\left(\partial_k \phi\right) \right]=A_8+2A_{11},
 & \partial_i \left[ \left(\Delta\phi\right)\left(\Delta \phi\right)
\left(\partial_i \phi\right) \right]=2A_9+A_{12},
 \\ \partial_i \left[\left(\Delta \phi\right)
\left(\partial_i \partial_j \phi\right)\left(\partial_j \phi\right) \right]
=A_9+A_{10}+A_{13},\qquad
 & \partial_i \left[ \left(\partial_i\partial_j\phi\right)
\left( \partial_j \partial_k \phi\right) \left(\partial_k \phi\right) \right]
=A_{10}+A_{11}+A_{14},
 \\ \partial_i \left[\left(\partial_j\partial_k \phi\right)
\left(\partial_j\partial_k \phi\right)\left(\partial_i \phi\right) \right]
=2A_{11}+A_{13}.
  \end{array}
\label{totalderivative}
\end{eqnarray}
Actually, two of the combinations in (\ref{totalderivative}) are not
independent and (\ref{totalderivative}) include only eleven independent
combinations. As a result, we can represent arbitrary marginal cubic
terms in the action as linear combinations of three among fourteen in
(\ref{A}).

Concretely, in this paper we shall use the following three terms. 
\begin{eqnarray}
S_3 = \int dt d^3x \, \frac{1}{M^5 a(t)^3} \left\{ \alpha_1 \phi^2 \Delta^3
  \phi + \alpha_2  (\Delta^2 \phi) (\partial_i \phi)^2
 + \alpha_3     (\Delta \phi)^3\right\}. 
\label{appendixS3}
\end{eqnarray}
An advantage of this representation is that contributions from these
three terms roughly correspond to local, equilateral and orthogonal
shape of bispectrum, respectively.

If we impose the shift symmetry for $\phi$ then the number of
independent interaction terms is reduced. With the shift symmetry, we
have to consider only the last eight terms in (\ref{A}) and only the
last seven combinations in (\ref{totalderivative}). One of the last
seven combinations in (\ref{totalderivative}) is not independent. As a
result, the marginal cubic terms in the action with the shift symmetry
can be written as (\ref{appendixS3}) with $\alpha_1=0$.

\subsection{Quartic terms}

Similarly, we can write down twenty quartic terms with six spatial
derivatives as follows,
\begin{eqnarray}
  \begin{array}{lll}
 B_{1}\equiv\left( \Delta^3 \phi \right) \phi^3 ,
& B_{2}\equiv
\left(\Delta^2 \partial_i \phi \right)
\left( \partial_i \phi \right) \phi^2  ,
&B_{3}\equiv
\left( \Delta^2 \phi \right) \left(\Delta \phi\right) \phi^2  , \\
B_{4}\equiv
\left(\Delta\partial_i \partial_j \phi \right) 
\left(\partial_i \partial_j\phi \right) \phi^2 ,
&B_{5}\equiv
\left( \Delta^2 \phi \right) \left(\partial_i\phi\right)
\left(\partial_i  \phi\right) \phi ,
&B_{6}\equiv
\left( \Delta \partial_i \partial_j \phi \right) 
\left( \partial_i\phi\right)\left(\partial_j \phi\right) \phi , \\
B_{7}\equiv
\left( \Delta \partial_i\phi \right)\left(\Delta \partial_i \phi\right)
 \phi^2  ,
&B_{8}\equiv
\left( \partial_i\partial_j\partial_k \phi \right) 
\left(\partial_i\partial_j\partial_k \phi\right) \phi^2 ,
&B_{9}\equiv
\left( \Delta \partial_i \phi \right) \left(\Delta\phi\right)
\left(\partial_i \phi \right) \phi , \\
B_{10}\equiv
\left( \Delta \partial_i\phi \right) 
\left(\partial_i \partial_j\phi\right)\left(\partial_j \phi\right) \phi ,
&B_{11}\equiv
\left( \partial_i\partial_j\partial_k\phi \right) 
\left(\partial_i \partial_j\phi\right)\left(\partial_k \phi\right) \phi ,
&B_{12}\equiv
\left( \Delta\phi \right)\left(\Delta \phi\right)
\left(\Delta \phi\right) \phi , \\
B_{13}\equiv
\left( \Delta \phi \right)\left(\partial_i\partial_j \phi\right)
\left(\partial_i\partial_j \phi\right) \phi ,
&B_{14}\equiv
\left( \partial_i\partial_j\phi \right)\left(\partial_j\partial_k \phi\right)
\left(\partial_k\partial_i \phi\right) \phi,
&B_{15}\equiv
\left( \Delta\partial_i\phi \right) \left(\partial_i\phi\right)
\left(\partial_j \phi\right)\left(\partial_j \phi\right) , \\
B_{16}\equiv
 \left( \partial_i\partial_j\partial_k\phi \right) 
\left(\partial_i\phi\right)\left(\partial_j \phi\right)\left(\partial_k \phi\right) ,
& B_{17}\equiv
\left( \Delta\phi \right)\left(\Delta \phi\right)
\left(\partial_i \phi\right)\left(\partial_i \phi\right),
& B_{18}\equiv
\left( \Delta\phi \right) \left(\partial_i \partial_j\phi\right)
\left(\partial_i \phi\right)\left(\partial_j \phi\right) , \\
B_{19}\equiv
\left( \partial_i\partial_j\phi \right)\left(\partial_i \partial_j \phi\right)
\left(\partial_k \phi\right)\left(\partial_k \phi\right) ,
&B_{20}\equiv
\left( \partial_i\partial_j\phi \right)\left(\partial_i\partial_k \phi\right)
\left(\partial_j \phi\right)\left(\partial_k \phi\right) .
& \\
  \end{array}
\label{B}
\end{eqnarray}
Apparently, there are sixteen total derivatives as follows, 
\begin{eqnarray}
  \begin{array}{ll}
\partial_i \left[
\left( \Delta^2 \partial_i \phi\right)\phi^3 
\right]=B_{1}+3B_{2},
& \partial_i \left[
\left(\Delta^2 \phi\right)\left(\partial_i\phi\right) \phi^2 
\right]=B_{2}+B_{3}+2B_{5}, \\
\partial_i \left[
\left(\Delta\partial_i\partial_j \phi\right)\left(\partial_j\phi\right) \phi^2 
\right]=B_{2}+B_{4}+2B_{6},
& \partial_i \left[
\left(\Delta\partial_i \phi\right)\left(\Delta\phi\right) \phi^2 
\right]=B_{3}+B_{7}+2B_{9}, \\
\partial_i \left[
\left(\Delta\partial_j \phi\right)\left(\partial_i\partial_j\phi\right) \phi^2 
\right]=B_{4}+B_{7}+2B_{10},
& \partial_i \left[
\left(\partial_i\partial_j\partial_k \phi\right)
\left(\partial_j\partial_k\phi\right)\phi^2 
\right]=B_{4}+B_{8}+2B_{11}, \\
\partial_i \left[
\left(\Delta\partial_i \phi\right)\left(\partial_j\phi\right)
\left(\partial_j \phi\right) \phi 
\right]=B_{5}+2B_{10}+B_{15},
& \partial_i \left[
\left(\Delta\partial_j \phi\right)\left(\partial_i\phi\right)
\left(\partial_j \phi\right) \phi 
\right]=B_{6}+B_{9}+B_{10}+B_{15}, \\
\partial_i \left[
\left(\partial_i\partial_j\partial_k \phi\right)
\left(\partial_j\phi\right)\left(\partial_k \phi\right) \phi 
\right]=B_{6}+2B_{11}+B_{16},
& \partial_i \left[
\left(\Delta \phi\right)\left(\Delta\phi\right)\left(\partial_i \phi\right) \phi 
\right]=2B_{9}+B_{12}+B_{17}, \\
\partial_i \left[
\left(\Delta \phi\right)\left(\partial_i \partial_j\phi\right)
\left( \partial_j\phi\right) \phi 
\right]=B_{9}+B_{10}+B_{13}+B_{18},
& \partial_i \left[
\left(\partial_i\partial_j \phi\right)
\left(\partial_j \partial_k\phi\right)\left(\partial_k \phi\right) \phi 
\right]=B_{10}+B_{11}+B_{14}+B_{20}, \\
\partial_i \left[
\left(\partial_j\partial_k \phi\right)
\left(\partial_j\partial_k\phi\right)\left(\partial_i \phi\right) \phi 
\right]=2B_{11}+B_{13}+B_{19},
& \partial_i \left[
\left(\Delta \phi\right)\left(\partial_i\phi\right)
\left(\partial_j \phi\right)\left(\partial_j \phi\right) 
\right]=B_{15}+B_{17}+2B_{18}, \\
\partial_i \left[
\left(\partial_j\partial_k \phi\right)\left(\partial_i\phi\right)
\left(\partial_j \phi\right)\left(\partial_k \phi\right) 
\right]=B_{16}+B_{18}+2B_{20},
& \partial_i \left[
\left(\partial_i\partial_j \phi\right)\left(\partial_j\phi\right)
\left(\partial_k \phi\right)\left(\partial_k \phi\right) 
\right]=B_{15}+B_{19}+2B_{20}. \\
  \end{array}
  \label{const.B}
\end{eqnarray} 
Two of them are not independent. Thus, marginal quartic terms in the
action can be written as a linear combination of six independent terms. 
In this paper, we use the following terms,
\begin{eqnarray}
&& S_4 = \int dt d^3x \, \frac{1}{M^6 a(t)^3} 
 \bigl\{ \beta_1 \phi^3 \Delta^3 \phi + \beta_2 \phi^2 (\Delta \phi)
  (\Delta^2 \phi) + \beta_3 \phi ( \Delta \phi)^3 \nonumber\\
&&\qquad\qquad\qquad\qquad\qquad\qquad\qquad
  + \beta_4 \phi^2
  (\Delta \partial_i \phi)^2 
  + \beta_5 \phi^2 (\partial_i \partial_j
  \partial_k \phi)^2 + \beta_6 (\partial_i \partial_j \partial_k \phi)
  (\partial_i \phi) (\partial_j \phi) (\partial_k \phi)\bigr\}. 
  \label{appendixS4}
\end{eqnarray}

With the shift symmetry, we have to consider only the last six terms
in (\ref{B}) and only the last three combinations in
(\ref{const.B}). Therefore, the number of the independent terms is
three. In the notation (\ref{appendixS4}), the shift symmetry imposes
the constraints as 
\begin{eqnarray}
\beta_1=\frac{1}{12}\beta_5, \qquad 
\beta_2=\beta_4+\frac{3}{4}\beta_5, \qquad 
\beta_3=-\beta_4-\frac{3}{2}\beta_5,
\end{eqnarray}
and then the action can be transformed into the following form, 
\begin{eqnarray}
S_4 = \int dt d^3x \, \frac{1}{M^6 a(t)^3} 
 \bigl\{
 (\beta_4+2 \beta_5)\left( \Delta\phi \right)^2
\left(\partial_i \phi\right)^2+
\beta_5 \left( \partial_i\partial_j\phi \right)^2
\left(\partial_k \phi\right)^2
+ \beta_6 (\partial_i \partial_j \partial_k \phi)
  (\partial_i \phi) (\partial_j \phi) (\partial_k \phi)\bigr\}.
\end{eqnarray}
This manifestly has the shift symmetry.


\end{document}